\documentclass[12pt]{article}
\usepackage{epsfig}
\usepackage[T1]{fontenc}
\usepackage[latin2]{inputenc}
\usepackage{colordvi}
\def\bea{\begin{eqnarray}}
\def\eea{\end{eqnarray}}
\def\bq{\begin{quote}}
\def\eq{\end{quote}}
\def\nn{\nonumber}

\def\gappeq{\mathrel{\rlap
{\raise.5ex\hbox{$>$}}
{\lower.5ex\hbox{$\sim$}}}}
\def\lappeq{\mathrel{\rlap{\raise.5ex\hbox{$<$}}
{\lower.5ex\hbox{$\sim$}}}}
\def\simlt{\stackrel{<}{{}_\sim}}
\def\simgt{\stackrel{>}{{}_\sim}}
\newcommand{\ynu}[1]{\mathbf{Y}_{\nu}^{#1}}

\newcommand{\beq}{\begin{equation}}
\newcommand{\eeq}{\end{equation}}
\newcommand{\OO}[1]{\mathbf{\Omega}_{#1}}

\newcommand{\Un}[1]{\mathbf{U}_\nu^{#1}} 

\def\tim{\tilde{m}_1}
\def\bmeg{BR(\mu\to e\gamma)}
\def\bteg{BR(\tau\to e\gamma)}
\def\btmg{BR(\tau\to \mu\gamma)}
\newcommand{\msl}[1]{\tilde{\mathbf{m}}_{L_{#1}}^2}
\def\mmaj{\mathbf{M}_R}
\def\arg{\mathrm{arg}\,}
\def\varkappa{\omega}
\def\gev{\,\mathrm{GeV}}

\newcounter{mnotecount}[section]

\begin{document}
\pagestyle{empty}
\begin{flushright}
{CERN-PH-TH/2004-030}\\
IFT-2004/15\\
{\tt hep-ph/0403180}\\
{\bf \today}
\end{flushright}
\vspace*{5mm}
\begin{center}

{\large {\bf Patterns of Lepton-Flavour Violation Motivated  by
Decoupling and Sneutrino Inflation}}\\
\vspace*{1cm}
{\bf Piotr~H.~Chankowski}$^1$, {\bf John~Ellis}$^2$, 
{\bf Stefan~Pokorski}$^1$, {\bf Martti~Raidal}$^{3}$ and {\bf 
Krzysztof~Turzy{\' n}ski}$^1$\\
\vspace{0.3cm}

$^1$  Institute of Theoretical Physics, Warsaw University, Ho\.za 69, 00-681, 
Warsaw, Poland\\
$^2$ Theoretical Physics Division, Physics Department, CERN, CH-1211 
Geneva 23, Switzerland 
\\
$^3$ National Institute of Chemical Physics and Biophysics,
Tallinn 10143, Estonia \\

\vspace*{1.7cm} 
{\bf Abstract} 
\end{center}
\vspace*{5mm}
\noindent
{
We present predictions for flavour-violating charged-lepton decays induced
by the seesaw mechanism implemented within the constrained minimal
supersymmetric standard model (CMSSM) with universal input soft
supersymmetry breaking terms. We assume that one heavy singlet neutrino
almost decouples from the see-saw mechanism, as suggested by the pattern
of light neutrino masses and mixing angles. This is suggested
independently by sneutrino inflation with a low reheating temperature,
$T_{RH}\simlt 10^7\,\mathrm{GeV}$, so as to avoid overproducing
gravitinos. This requirement further fixes the mass of the weakly-coupled
sneutrino, whose decays may lead to leptogenesis. We find that
$\bmeg\simgt 10^{-13}$ but $\btmg\simlt 10^{-9}$ in the bulk of the
acceptable parameter space, apart from a few isolated points. The ratio
$\bmeg/\bteg$ depends on only one complex parameter, and is particularly
interesting to compare with experiment.
}
\vspace*{1.0cm}
\date{\today} 


\vspace*{0.2cm}

\vfill\eject
\newpage

\setcounter{page}{1}
\pagestyle{plain}

\section{Introduction}

The observation of neutrino mixing \cite{mnuexp} has led many authors to 
consider the possibility of flavour violation and CP violation in the 
charged-lepton sector \cite{BOMA}. The most predictive framework for such 
studies is the seesaw model for neutrino masses~\cite{GERASL} implemented 
within the minimal supersymmetric extension of the Standard Model, 
constrained to have universal soft supersymmetry breaking terms at some input 
renormalization scale characteristic of grand unification (CMSSM). This 
framework allows one to explore the possible links between lepton-flavour 
violation (LFV), neutrino oscillations and leptogenesis \cite{FUYA,BUDIPL}, 
assuming 
the cosmological baryon asymmetry to have originated from CP-violating decays 
of heavy singlet neutrinos and their supersymmetric partners.

Even the minimal three-generation seesaw model contains 18 free parameters
\cite{CAIB, DAIB}, in addition to those describing supersymmetry breaking. 
Four combinations of these seesaw parameters have been determined by 
neutrino-oscillation experiments, two mixing angles and two mass-squared 
differences. More of the seesaw parameters may be determined by low-energy 
neutrino experiments, but not all of them, and specifically not all those 
controlling leptogenesis \cite{various,ELRA,DA,DAIB3,BR}. Under these circumstances, 
supplementary
hypotheses are needed if one is to make unambiguous predictions for LFV,
leptonic CP violation and leptogenesis. Fortunately, one can find good 
physical motivation for certain simplifying assumptions. The observed 
hierarchy of differences in neutrino masses-squared can be interpreted as 
a hierarchy in the masses themselves. Such a pattern of the active 
neutrino 
masses together with the measured nearly bi-maximal neutrino mixing would 
be naturally explained if one of the heavy singlet neutrinos is almost 
decoupled from the seesaw mechanism \cite{KI,ALFE}. As 
we analyze in this paper, this decoupling hypothesis imposes important 
constraints on the seesaw parameters, and hence leads to interesting 
predictions for LFV processes.

The decoupling hypothesis is supported by the suggestion that the scalar
field supposed to be responsible for inflation, the inflaton, is one of
the heavy singlet sneutrinos~\cite{SN}. Requiring a low reheating 
temperature, $T_{RH}\simlt 10^7$~GeV after inflation, in order to solve the 
cosmological gravitino problem \cite{GRAVIT}, forces the inflaton sneutrino 
to couple very weakly to ordinary matter and its superpartner almost to
decouple from the seesaw mechanism. In order to explain simultaneously the 
duration of the inflationary epoch responsible for the large-scale structures 
observed in the Universe and the magnitudes of the perturbations observed in 
the cosmic microwave background, the mass of this sneutrino should be around 
$2 \times 10^{13}$~GeV~\cite{ELRAYA2}. This is well inside the range 
thought 
plausible in the seesaw model. Also, as discussed elsewhere, sneutrino 
inflation makes predictions for cosmic microwave background (CMB) observables, 
such as the scalar spectral index and the ratio of tensor to scalar 
perturbations~\cite{ELRAYA2}, which are consistent with data on the CMB and 
cosmological structure formation~\cite{WMAP}. 

It is striking that, on the one hand, naturalness arguments for models of
neutrino masses and mixings and, on the other hand, the hypothesis of the
sneutrino inflation, both motivate independently a similar pattern in the
seesaw mechanism.

The scenario with the heaviest right-chiral neutrino decoupled is a
natural possibility. However, a hierarchy of the heavy singlet neutrino
masses could be compensated by some hierarchy in the neutrino Yukawa
couplings, and it is interesting to consider also the decoupling of one of
the lighter neutrinos. In fact, decoupling of the lightest singlet
neutrino was considered in the sneutrino inflation model of~\cite{ELRAYA2},
but one may also consider inflation driven by one of the heavier
sneutrinos, if it is decoupled. The purpose of this paper is to outline
the different scenarios for decoupling one of the heavy singlet neutrinos,
and explore in some detail the LFV signature of some specific examples.

We discuss our different decoupling assumptions for the neutrino Yukawa
couplings in Section~\ref{sectwo}, and explore their various predictions
for LFV decays in Section~\ref{seclfv}. For each case we also discuss
the possible leptogenesis, which is necessarily nonthermal in the 
sneutrino inflation models with low reheating temperature,
with the CP asymmetry generated in the decays of the inflaton \cite{LASH}. 
As we
discuss in Section~\ref{seclfv}, the details of this mechanism depend on
which neutrino plays the role of the inflaton.

The main result of our investigations is that the decoupling hypothesis
leads to relatively rigid predictions for the branching ratios
$BR(\ell_i\to\ell_j\gamma)$. For most of the acceptable parameter range, 
we predict $\bmeg\simgt 10^{-13}$,
within the reach of the experiment now underway at PSI. However, for
generic models with $\bmeg$ below the present experimental upper limit, we
find that $\btmg < 10^{-9}$, apart from very particular parameter points,
which may be of special theoretical interest. Only in these cases might
$\btmg$ be within the reach of present experiments at $B$ factories and
the LHC.  The ratios $BR(\ell_i\to\ell_j\gamma)/BR(\ell_k\to\ell_m\gamma)$
depend, within our assumptions, on only one unknown complex parameter that
appears in the neutrino Yukawa couplings. If possible, a measurement of
$\btmg / \bmeg$ would provide particularly interesting information on
complex structure in the seesaw model.

\section{Decoupling Assumptions for Yukawa Couplings}
\label{sectwo}

In the seesaw mechanism the observed neutrino masses and mixing are
determined by the neutrino Yukawa couplings $\ynu{AB}$ and the mass matrix 
$\mmaj$ of the heavy singlet neutrinos. The mass 
matrix of the light left-chiral neutrinos is given by:
\beq
\label{numasses}
(\mathbf{m}_\nu)_{AB} = \langle H \rangle^2 \mathbf{C}_{AB}
\eeq
where $\langle H\rangle$ is the electroweak symmetry breaking vacuum 
expectation value of the relevant Higgs doublet and the complex symmetric 
matrix $\mathbf{C}$ is the coefficient of the dimension-five operator
$\mathcal{L}_\mathrm{eff}=-\mathbf{C}_{AB}(L_AH)(L_BH)/2$  resulting from 
integrating out heavy right-chiral neutrinos \cite{GERASL}. The 
matrix $\mathbf{C}$, which can be diagonalized by a unitary 
matrix $\mathbf{U}_\nu^T\mathbf{C}\Un{}=\mathrm{diag}(c_1^2,c_2^2,c_3^2)$,
is at the high scale given by
\beq
\label{seesaw}
\mathbf{C} = -\ynu{T}\mmaj^{-1}\ynu{}~.
\eeq
Similarly as for the quark sector, the number of observables in the light 
neutrino sector is smaller than the number of free parameters in $\ynu{}$ 
and $\mmaj$. Indeed, even in the basis in which 
${\mmaj}=\mathrm{diag}(M_1,M_2,M_3)$ and 
$\mathbf{Y}_e=\mathrm{diag}(y_e,y_\mu,y_\tau)$, the most general solution 
of the relation (\ref{seesaw}) for $\ynu{}$ reads:
\beq
\ynu{AB} = i M_A^{1/2}\sum_C \OO{AD}c_D {\Un{BD}}^\ast~.\label{YukawaOmega}
\eeq
A complex orthogonal matrix $\OO{}$ 
accounts for the six-parameter ambiguity in translating the 3 neutrino 
masses and 6 parameters in $\Un{}$ into the 15 parameters of the neutrino 
Yukawa coupling $\ynu{}$ \cite{CAIB}. Thus, even assuming that the 
elements of the mixing matrix $\mathbf{U}_\nu$ are determined with sufficient 
accuracy, the predictions of the CMSSM for flavour-violating processes
in the lepton sector 
depend on several unknown factors: the pattern of the light neutrino masses 
(hierarchical, inversely hierarchical or degenerate), the right-chiral 
neutrino masses $M_A$ and on the matrix $\OO{}$. In practice, one should 
also remember that the Majorana phases in $\Un{}$ may never be measured.

In a large number of papers, the predictions for the charged lepton decays 
have been discussed under various specific assumptions about those unknown 
factors 
\cite{CAIB,DAIB,various,ELRA,DA,HIMOTOYAYA,BIFEZH,BLKI,
ELHILORA,LAMASA,PEPRTAYA,IBRO2,TWY,MPVY,CMM}. 
In the present paper we re-examine the predictions for the charged lepton 
decays of the CMSSM under two very general assumptions. We assume the 
hierarchy in mass for both, left- and right-chiral neutrinos
\begin{eqnarray}
&m_{\nu_1} \ll m_{\nu_2}\leq m_{\nu_3}& \nonumber \\
& M_1 \leq M_2 < M_3 &\label{assumpt}
\end{eqnarray}
We take $m_{\nu_3}=\sqrt{\Delta m_\mathrm{atm}^2}=0.05$~eV,  
$m_{\nu_2}=\sqrt{\Delta m_\mathrm{sol}^2}=0.008$~eV and 
the neutrino mixing matrix in the form: 
\begin{eqnarray}
\mathbf{U}_\nu=\left(\matrix{
c_{12} & s_{12} & s_{13}e^{-i\delta}\cr
-\frac{s_{12}}{\sqrt{2}}+\dots & 
\frac{c_{12}}{\sqrt{2}}+\dots & {1\over\sqrt2} \cr
{s_{12}\over\sqrt{2}}+\dots & -{c_{12}\over\sqrt{2}}+\dots & 
{1\over\sqrt2}} \right)
\cdot\mathrm{diag}(e^{i\phi_1},e^{i\phi_2},1)
\end{eqnarray}
where $s_{12}^2\equiv\sin^2\theta_{12}=0.315$, the dots stand for small
terms $\sim{\cal O}(s_{13})$ (for the sake of definiteness we have
assumed that $0<\theta_{23}<\pi/2$). For the
$\mathbf{U}_\nu^{13}$ entry we use the two scenarios specified in Table 1.
We do not assume any particular 
values of the Majorana phases $\phi_1$ and $\phi_2$, but treat them as free
parameters. It is important to note that, for hierarchical light neutrino
masses, renormalization effects on the mass eigenvalues and on the mixing
angles, in particular, are small and can be neglected 
(see e.g. \cite{CHKRPO,CHPO}). 
Thus one can use the measured $\Un{}$ in (\ref{YukawaOmega}). 

\begin{table}
\begin{center}
\begin{tabular}{|c|cc|}
\hline
case & $|\Un{13}|$ & $\arg \Un{13}$ \\
\hline
a) & 0 & $-$ \\
b) & 0.1 & $-\pi/2$ \\
\hline
\end{tabular}
\end{center}
\caption{\it Selected values of $\Un{13}$ at the electroweak scale used in 
the analysis.\label{table1}}
\end{table}

It has frequently been observed that the simultaneous appearance of 
hierarchical light neutrino masses and two large mixing angles is not 
'natural' in the seesaw mechanism (see e.g. \cite{ALFE}). Important 
exceptions are so-called sequential neutrino dominance models with one 
neutrino decoupled \cite{KI,LAMASA} or even absent \cite{FRGLYA,BHR,IBRO2,KUMO}. Otherwise 
some particularities, as found, e.g., in models with horizontal 
symmetries~\cite{ALFE2,GRNISH,HIKI}, are necessary. This motivates our 
second assumption 
that one singlet 
right-chiral neutrino, not necessarily the heaviest one, decouples from 
the seesaw mechanism, in the sense that at most one neutrino, say $N_A$, 
contributes to the mass $m_{\nu_1}$ of the lightest neutrino. As has been 
explained in~\cite{CHTU,IBRO1}, for $\mathbf{Y}_\nu^{A2}/M_A\rightarrow0$ and 
$\mathbf{Y}_\nu^{A3}/M_A\rightarrow0$ the matrix $\OO{}$ in 
(\ref{YukawaOmega}) has to be, depending on the index $A$, of one of the 
following three forms:
\begin{eqnarray}
\label{omegalep}
\textrm{decoupling of $N_1$} & & \OO{} = 
\left(\begin{array}{ccc} 1 & 0 & 0 \\ 0 & z & p \\ 0 & \mp p & \pm z 
\end{array}\right) \\
\label{omegatwo}
\textrm{decoupling of $N_2$} & & \OO{} = 
\left( \begin{array}{ccc} 0 & z & p \\ 1 & 0 & 0 \\ 0 & \mp p & \pm z 
\end{array}\right) \\
\label{omegadec}
\textrm{decoupling of $N_3$} & & \OO{} = 
\left( \begin{array}{ccc} 0 & z & p \\ 0 & \pm p & \mp z \\ 1 & 0 & 0 
\end{array}\right)
\end{eqnarray}
where $z^2+p^2=1$. Since $\OO{AB}^2$ determines directly the contribution of 
the right-chiral neutrino $N_A$ to the mass of the light neutrino $\nu_B$ 
\cite{LAMASA}, the assumption that $N_A$ effectively decouples means 
that it is only $N_A$ that contributes to $m_{\nu_1}$, and that it does 
not contribute to $m_{\nu_{2,3}}$. In the limit of strict decoupling, in 
which $\mathbf{Y}_\nu^{AB}/M_A\rightarrow0$ for $B=1,2,3$, the lightest 
left-chiral neutrino is massless because, as follows from 
(\ref{omegalep})-(\ref{omegadec}) and (\ref{YukawaOmega}),
$\sqrt{m_{\nu_1}}\propto (\mathbf{Y}_\nu\mathbf{U}_\nu)^{A1}/\sqrt{M_A}$.
Of course, in realistic cases the zeroes in $\OO{}$ are non-zero numbers 
$\ll1$.
The stability of the patterns (\ref{omegalep})-(\ref{omegadec}) with 
respect to radiative corrections is analyzed in the Appendix.

Another motivation for seesaw models with the light-neutrino masses
dominated by two heavy singlet neutrinos follows from the hypothesis
that the cosmological inflaton field is one of the heavy sneutrinos,
$\tilde{N}_A$, say. As has been discussed in \cite{ELRAYA2}, 
in order to reproduce the measured characteristics of the CMB in such a 
scenario,
and to agree with the data on the cosmological structure formation, 
the inflaton-sneutrino and its superpartner, the right-chiral neutrino 
$N_A$ must have a mass $M_A \simeq 2\times 10^{13}\gev$. The reheating 
temperature following inflaton decay is
\beq
T_{RH} \sim \sqrt{\tilde m_A M_P}{M_A \over\langle H\rangle},\label{eqn:trh}
\eeq
where 
\beq
\tilde m_A\equiv\left(\ynu{}\ynu{\dagger}\right)_{AA}\langle H\rangle^2/M_A 
= \sum_{B=1}^3|\OO{AB}|^2 m_{\nu_B}~. 
\eeq
For $M_A \simeq 2\times 10^{13}\gev$, it follows from (\ref{eqn:trh}) that
one needs $\tilde m_A\simlt 10^{-17}$ eV in order to obtain $T_{RH}\simlt10^7$ 
GeV \cite{TRHbound}. This in turn enforces a pattern in the $\OO{}$ matrix 
with $(1,0,0)$ in 
the $A$'th row. Thus reheating temperature $T_{RH}$ after inflation that 
is low enough to solve the cosmological gravitino problem is possible in 
sneutrino 
inflation scenarios provided the neutrino partner of the $\tilde{N}_A$ 
inflaton (not necessarily the heaviest one) decouples from the seesaw 
mechanism.
The possibility of explaining the baryon asymmetry of the Universe 
through leptogenesis following inflation with the patterns 
(\ref{omegalep})-(\ref{omegadec})
will be discussed in Sec. \ref{seclfv}.

The three patterns of $\OO{}$ shown in (\ref{omegalep}), (\ref{omegatwo}) 
and 
(\ref{omegadec}) provide distinctly different realizations of the seesaw 
mechanism. However, as we explain later, they give similar predictions 
for LFV decays in a large range of parameter space. It follows that the 
predictions for neutrino masses and mixings and for LFV are invariant under 
certain transformations on the neutrino Yukawa matrix. Therefore, even with 
all possible data on neutrino masses and mixings, as well as on LFV processes 
that can (in principle) be obtained experimentally, the bottom-up approach 
can determine $\ynu{}$ only up to these transformations.

The analysis presented in this paper is based only on the general assumptions 
described above. It is, however, worth commenting on the link of such a 
bottom-up approach to model-dependent (top-down) studies based on specific 
`theories' of the  neutrino Yukawa matrix and of the masses of the heavy 
singlet neutrinos, based for example on texture zeroes. Once such a theory of 
flavour is specified in some electroweak basis, as in~\cite{IBRO1,IBRO2} for 
example, $\OO{}$ can be read off easily from (\ref{YukawaOmega}),
by first performing the necessary field rotations diagonalizing the 
heavy singlet neutrino mass matrix and the charged lepton Yukawa coupling 
matrix. One may envisage a situation in which the 
electroweak basis in which the flavour theory predicts texture zeroes in 
$\ynu{}$ is only slightly different from the basis in which our assumptions
are formulated, i.e., the matrices $\mathbf{Y}_e$ and $\mathbf{M_R}$ 
are almost 
diagonal. The equation (\ref{YukawaOmega}) fixes then the value of the 
parameter $z$ in terms of the ratios of some elements of $\Un{}$ and of
$m_{\nu_2}/m_{\nu_3}$. Indeed, for the pattern (\ref{omegalep}) [pattern 
(\ref{omegatwo}), (\ref{omegadec})], $\ynu{1A}$ [$\ynu{2A}$, $\ynu{3A}$]
can be neglected, and if in addition the texture gives $\ynu{2B}=0$
[$\ynu{1B}=0$] for $B=2,3$ then one obtains from (\ref{YukawaOmega}) 
\beq
\label{y22tz}
\pm z = 1-\frac{1}{2}{m_{\nu_2}\over m_{\nu_3}}
\left({\Un{B2*}\over\Un{B3*}}\right)^2\approx 1\mp 0.06 e^{-2i\phi_2}~.
\eeq
Similarly, if $\ynu{3B}=0$ [$\ynu{3B}=0$, $\ynu{2B}=0$] for $B=2,3$ 
because of texture zeroes, then
\beq
\label{y32tz}
z = \pm \sqrt{\frac{m_{\nu_2}}{m_{\nu_3}}}\frac{\Un{B2*}}{\Un{B3*}} 
\approx \pm 0.3 e^{-i\phi_2}~.
\eeq
For a diagonal matrix $\mathbf{M}_R$,
the solutions (\ref{y32tz}) and (\ref{y22tz}) remain approximately valid if
\beq
\label{irbound}
|\mathbf{U}_\ell^{AB}|\simlt \sqrt{\frac{m_{\ell_A}}{m_{\ell_B}}} 
\phantom{aa} {\rm for} \phantom{aa} A < B,
\eeq
where $\mathbf{U}_\ell$ is the matrix diagonalizing 
$\mathbf{Y}_e^\dagger \mathbf{Y}_e$. The solution (\ref{y22tz}) is more 
sensitive to departures of $\mathbf{M}_R$ from the diagonal form than is
(\ref{y32tz}). It will be useful to remember these particular qualitative 
patterns with $z\approx 0.3$ or $z\approx 1$ in our subsequent analysis.


\section{Implications of Decoupling for LFV Decays}
\label{seclfv}

In this Section we first present the general background for calculating the 
branching ratios for the LFV decays $l_A\to l_B\gamma$ and then discuss in 
detail the predictions that follow from the pattern (\ref{omegalep}) of the 
matrix $\OO{}$. The predictions following from the two other patterns are
similar and will require only a short discussion. For each of the patterns
we also discuss possible scenarios for leptogenesis.

\subsection{General Formalism}

The branching ratios for LFV decays are well described by a 
single-mass-insertion approximation~\cite{HIMOTOYAYA,LAMASA}:
\beq
\label{brdef}
BR(l_A\to l_B\gamma)\approx{\alpha^3\over G_F^2} 
\mathcal{F}(m_0,M_{1/2},\mu)|\msl{AB}|^2 \tan^2\beta,
\eeq
where $\mathcal{F}$ is a function of the soft supersymmetry-breaking 
masses fixed at the high scale $M_X$. The 
ratio $f(m_0,M_{1/2}) \equiv \mathcal{F}(m_0,M_{1/2}) / 
\mathcal{F}(100~{\rm GeV}, 500~{\rm GeV})$ for $\tan \beta = 10$ and $\mu 
> 0$ is shown Fig.~\ref{fiffun}a. 
The reference values $m_0 = 100~{\rm GeV}, M_{1/2} = 500~{\rm GeV}$ 
are chosen for their consistency with the astrophysical cold dark matter 
constraint~\cite{ELOLSASP}. In the subsequent figures, we show predictions 
with the factor $f(m_0,M_{1/2})$ removed, and  Fig.~\ref{fiffun}a can be 
used to obtain the corresponding predictions for other values of 
$m_0, M_{1/2}$.

The 
off-diagonal entries in the slepton mass matrix $\msl{}$ are generated 
radiatively by the renormalization-group (RG) evolution from $M_X$ to the 
electroweak scale. In the commonly-used approximation of using only a single 
recurrence in solving the relevant RGEs, they are related to the neutrino 
Yukawa couplings by
\beq
\label{RGEsol1}
\msl{AB} = \kappa \sum_C (\ynu{CA})^* (\Delta t+\Delta\ell_C) \ynu{CB},
\eeq
where $\kappa =-6m_0^2-2A_0^2$, $\Delta t=\ln(M_X/M_3)/16\pi^2$,
$\Delta\ell_C=\ln(M_3/M_C)/16\pi^2$ \cite{ELHIRASH}. This result is not 
very accurate, as we demonstrate in Fig.~\ref{fiffun}b, where we show 
the ratio of (\ref{RGEsol1}) to $\msl{AB}$ as
obtained by solving numerically the one-loop RGEs.
Although the results presented in this paper are based on the exact
numerical integration  of the RGEs with successive decoupling of
the singlet neutrinos at the appropriate scales, we note that
the numerical result is well approximated by the following
correction to $\kappa$:
\begin{eqnarray}
\Delta\kappa=\left[-\frac{36}{5}g^2(2M_{1/2}^2+3m_0^2 +A_0^2+2A_0M_{1/2}) 
\right.\phantom{aaaaaaaa} \nonumber\\
\left.+ y_t^2(36m_0^2+24A_0^2) +(6m_0^2+12A_0^2)
\mathsf{Tr}(\ynu{\dagger}\ynu{}) \right] \Delta t,\label{eqn:dkappa}
\end{eqnarray}
where the top-quark Yukawa $y_t$ and the other couplings are taken at 
the scale $M_X$. The correction  (\ref{eqn:dkappa}) is obtained by 
including the second recurrence in solving the RGEs. However, the 
approximation (\ref{RGEsol1}) is sufficient for the qualitative discussion 
of our results.

Combining (\ref{RGEsol1}) with (\ref{YukawaOmega}) and 
(\ref{assumpt}) and using one of the decoupling scenarios 
(\ref{omegalep})-(\ref{omegadec}), one may obtain formulae for the decay 
rates
$BR(l_A\to l_B\gamma)$. It is useful to discuss first the results
obtained under the technical assumption that
the hierarchies of neutrino masses obey
\begin{eqnarray}
m_{\nu_3} M_2 < m_{\nu_2} M_3 &\qquad & 
\textrm{for pattern (\ref{omegalep}),} \nonumber \\
m_{\nu_3} M_1 < m_{\nu_2} M_3 &\qquad & 
\textrm{for pattern (\ref{omegatwo}), or} \label{assumpt2}\\
m_{\nu_3} M_1 < m_{\nu_2} M_2 &\qquad & 
\textrm{for pattern (\ref{omegadec}),}\nonumber
\end{eqnarray}
where the inequality signifies a ratio of at least a factor 2 or 3.
We study deviations from  (\ref{assumpt2}) at the end of Subsection 
\ref{secN1}. 
As follows from eq. (\ref{YukawaOmega}), under these assumptions and
for generic values of the $z$ and $p$ parameters, the formula (\ref{RGEsol1}) 
is well approximated by the contribution of the product
$(\ynu{3A})^\ast\ynu{3B}$ for the structures (\ref{omegalep})
and (\ref{omegatwo}), and of the product
$(\ynu{2A})^\ast\ynu{2B}$ for the structure (\ref{omegadec}). Nevertheless,
for the sake of the future discussion, in the formulae given below we
display also the nonleading contribution to the formula (\ref{RGEsol1}). 
In our numerical calculation we keep, of course, complete $\msl{AB}$
obtained by solving the RGEs.

\subsection{The Decoupling of $N_1$ and LFV Decays}
\label{secN1}

With the pattern (\ref{omegalep}), 
the mass insertion $\msl{32}$ relevant for $\tau\to\mu\gamma$ reads:
\bea
\label{q32}
\msl{32} & \approx &
\kappa\frac{m_{\nu_3}M_3\Delta t}{\langle H\rangle^2}\times\\ 
&&\left[\Un{33}\Un{23\ast}\left(|z|^2 + S |1-z^2|\right) +
R \,\Un{33}\Un{22\ast}
\left(S z\sqrt{1-z^2}^\ast - z^\ast\sqrt{1-z^2}\right) + \right.
\nonumber\\ 
&& \left.
+R\,\Un{32}\Un{23\ast}\left(S z^\ast\sqrt{1-z^2} - z\sqrt{1-z^2}^\ast\right) +
R^2\,\Un{32}\Un{22\ast}\left( S |z|^2 + |1-z^2| \right) 
\right], \nn
\eea
where $R=\sqrt{m_{\nu_2}/m_{\nu_3}}\sim 0.41$ and 
$S=M_2(1+\Delta\ell_2/\Delta t)/M_3\sim M_2/M_3$ gives the subleading 
contribution of the product $(\ynu{2A})^\ast\ynu{2B}$. 
The branching ratio does not depend strongly on the masses $M_1$ 
and $M_2$ and the Majorana phase $\phi_1$, as long as $M_1<M_2\ll M_3$, 
i.e., for $S\ll1$. For the results presented below, we take
$M_3=5\times10^{14}$~GeV, $M_2=3\times 10^{13}$~GeV ($S\approx 0.1$) and 
$M_1=2\times 10^{13}$~GeV, consistent with inflation being driven by the 
lightest singlet sneutrino. Our choice of $M_3$ makes $\ynu{3A}$ naturally 
of order of unity. Moreover, due to maximal atmospheric neutrino mixing, 
$\mathsf{Re}(\Un{33}\Un{22\ast}-\Un{32}\Un{23\ast})\sim\mathcal{O}(s_{13})$ 
and the real parts of the second and third term in (\ref{q32}) cancel each 
other almost completely. Then it is only $\mathsf{Im}(\,\msl{32})$ which 
depends on the Majorana phase $\phi_2$, and the branching ratio 
$BR(\tau\to\mu\gamma)\propto|\msl{32}|^2$ consist of a sum of two positive 
terms: a constant one and a term which oscillates with $\phi_2$ as 
$\sin^2[\phi_2-\arg(z^\ast\sqrt{1-z^2})]$. The relative magnitude of these 
two terms depends on the phase of $z$, and a non-zero value of $\Un{13}$ 
adds a slight modulation proportional to 
$-\sin(\phi_2-\arg(z^\ast\sqrt{1-z^2}))$. 

\begin{figure}
\includegraphics*[height=8cm]{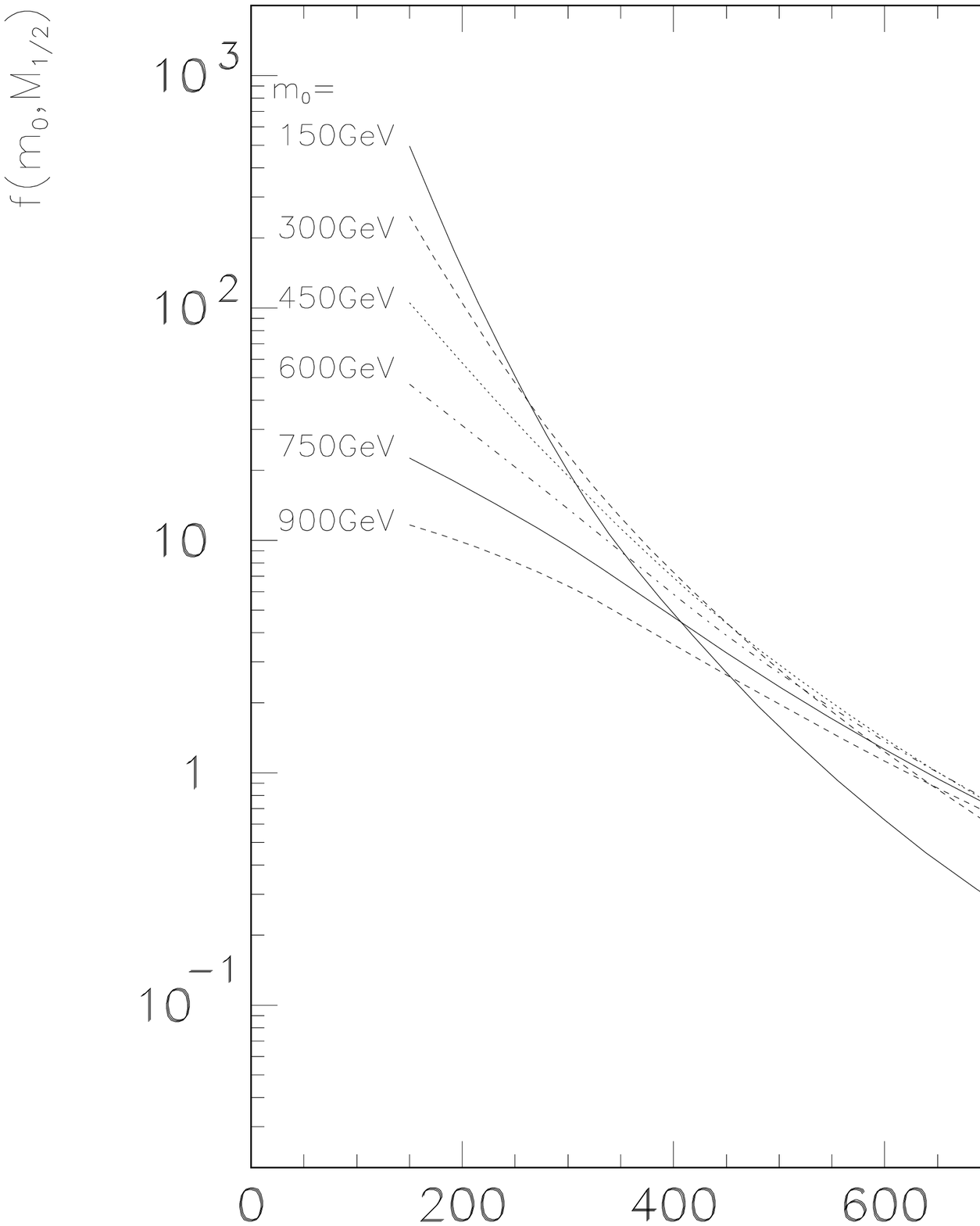}
\hspace{1cm}
\includegraphics*[height=8cm]{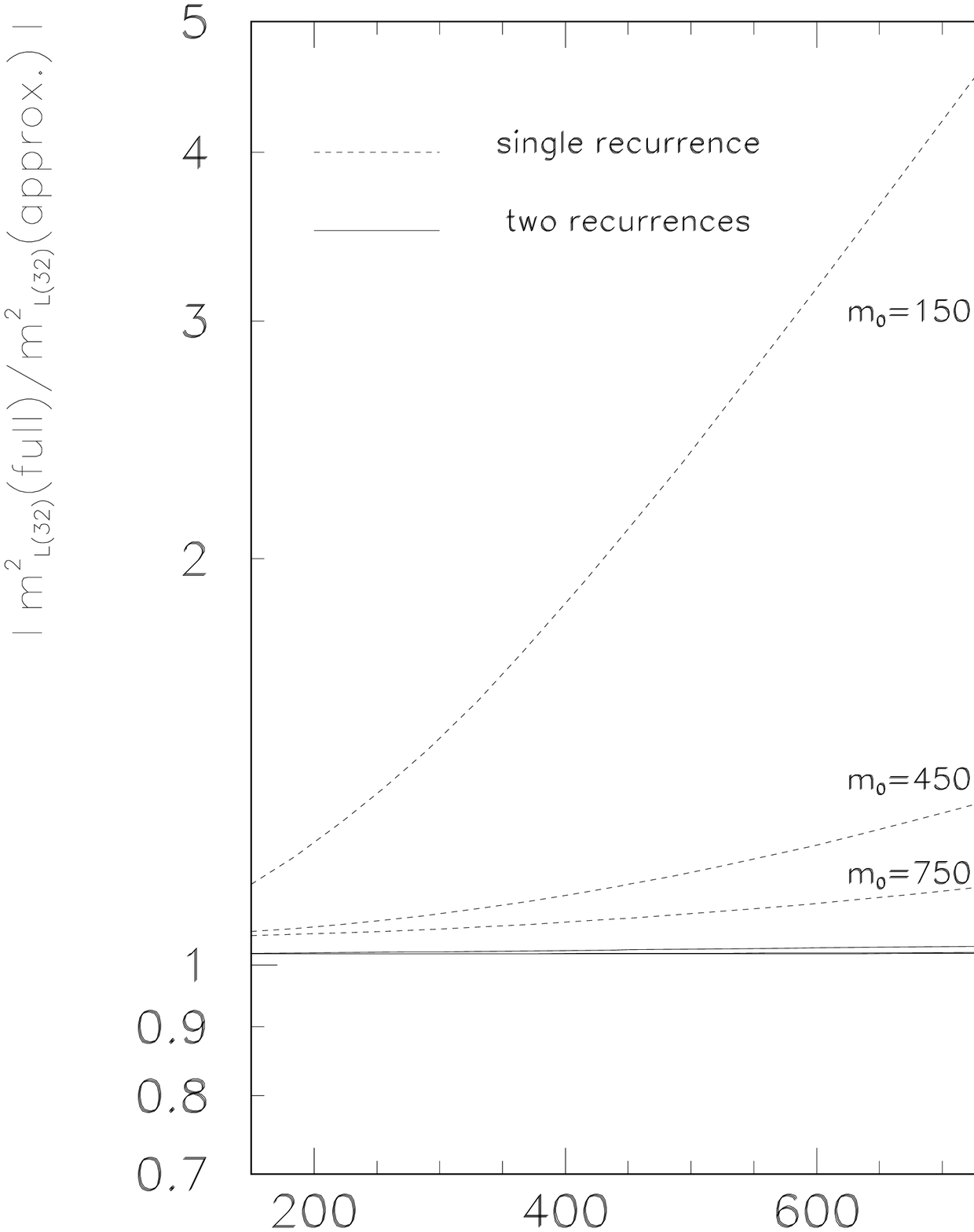}
\caption{\it Left panel: $f(m_0,M_{1/2})\equiv\mathcal{F}(m_0,M_{1/2})/
\mathcal{F}(100\gev,500\gev)$ as a function of $M_{1/2}$ for 
$\tan\beta=10$ and $\mu>0$. The normalization is such that
$f(m_0,M_{1/2})=1$ for $m_0=100$~GeV, $M_{1/2}=500$~GeV, consistent with 
the dark matter constraints~\cite{ELOLSASP}. Right panel: The ratio of 
${\msl{32}}(\mathsf{full})$ obtained by numerical integration of the full 
set of the RGEs to ${\msl{32}}(\mathsf{approx.})$ obtained with the use 
of the formula (\ref{RGEsol1}) as a function of $M_{1/2}$ for $\tan\beta=10$ 
and $A_0=0$. \label{fiffun}}
\end{figure}

\begin{figure}
\includegraphics*[height=8cm]{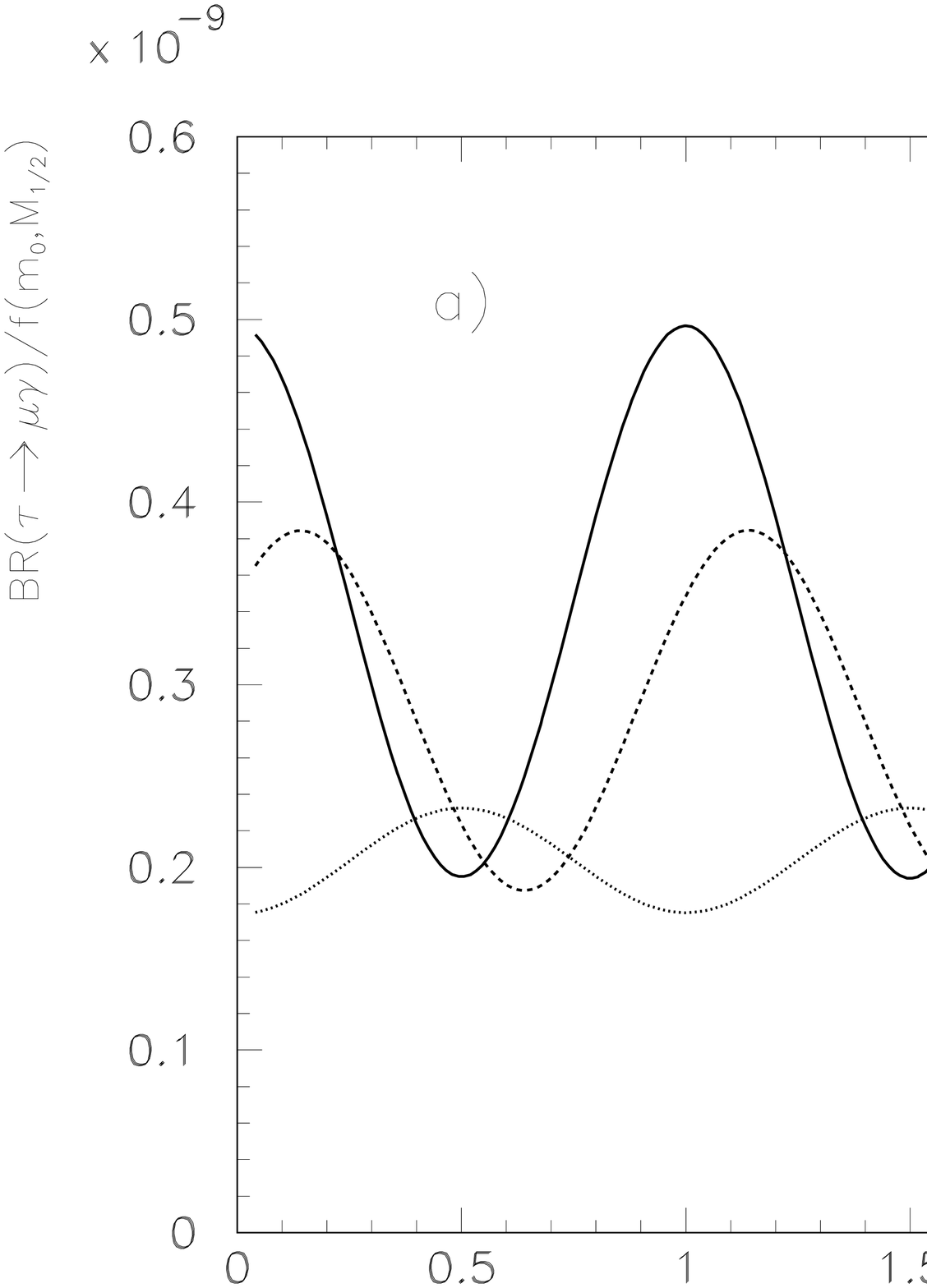}
\hspace{1cm}
\includegraphics*[height=8cm]{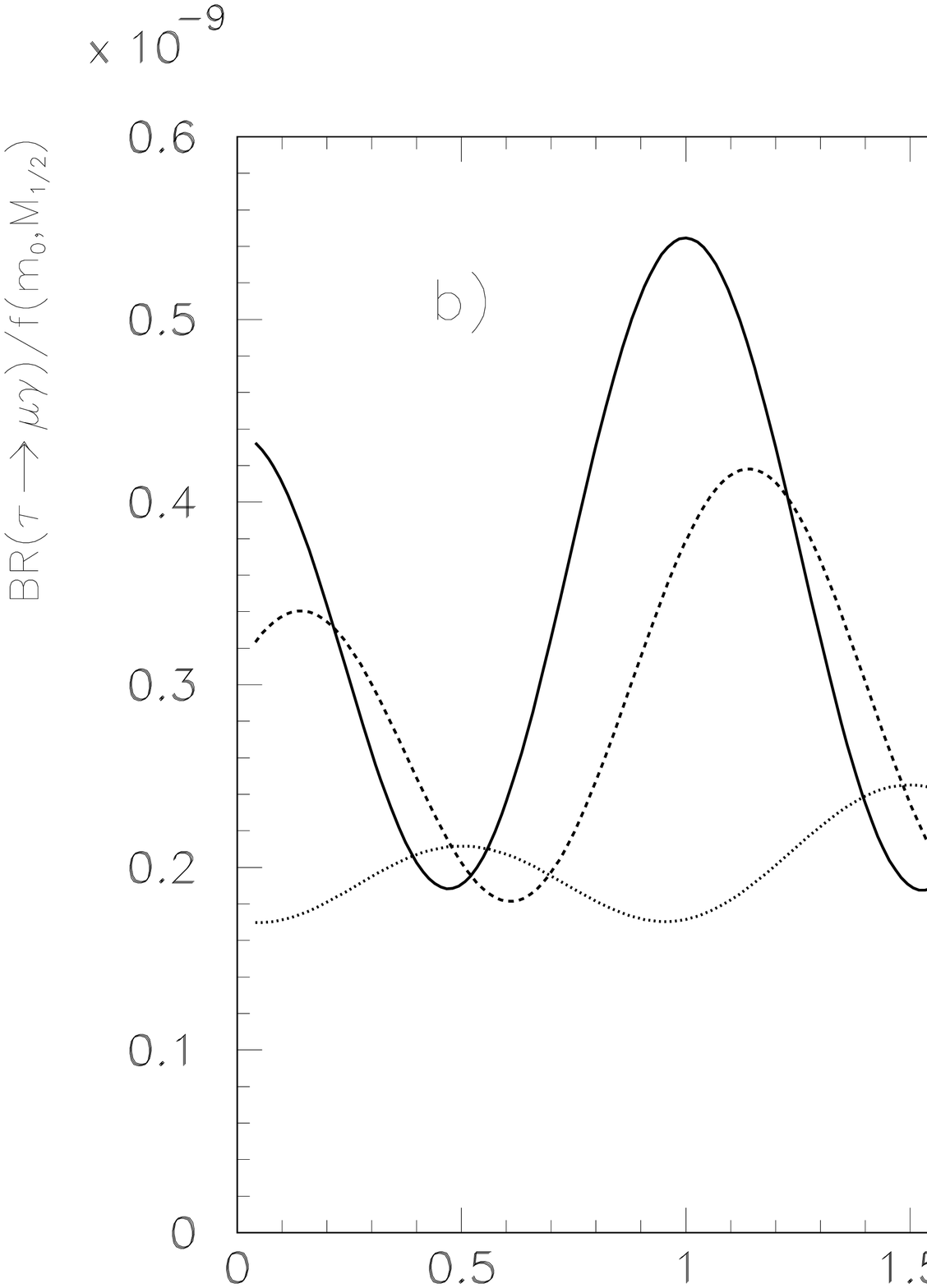}
\caption{\it $BR(\tau\to\mu\gamma)/f(m_0,M_{1/2})$ as a function of 
$\phi_2$ 
for the choice of the remaining phases described in Table \ref{table1}, 
$|z|=1/\sqrt2$, $\tan\beta=10$ and $A_0=0$. Dotted, dashed and solid lines 
correspond to $\arg z=0,\pi/4,\pi/2$, respectively. \label{fitmg1}}
\end{figure}

\begin{figure}
\includegraphics*[height=8cm]{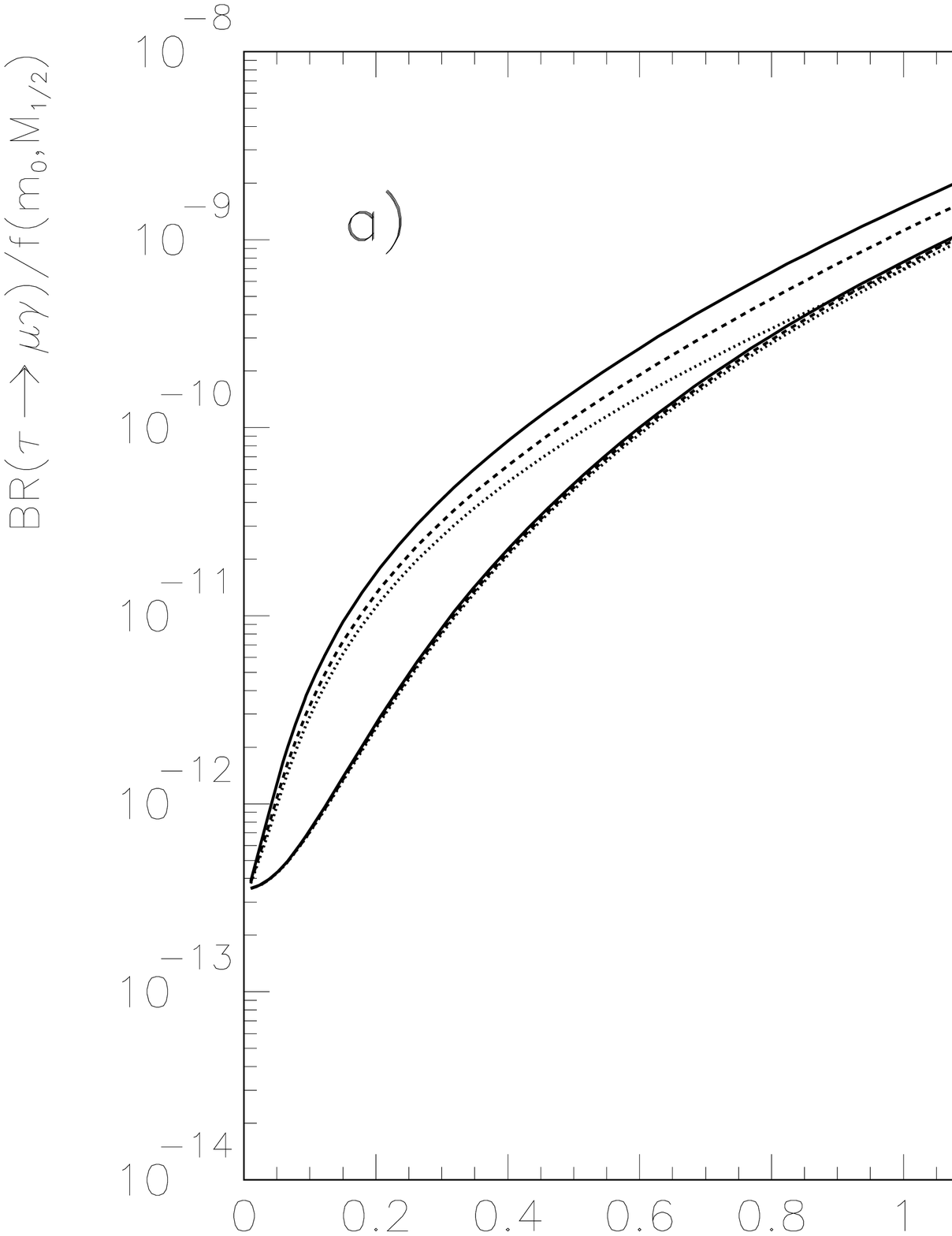}
\hspace{1cm}
\includegraphics*[height=8cm]{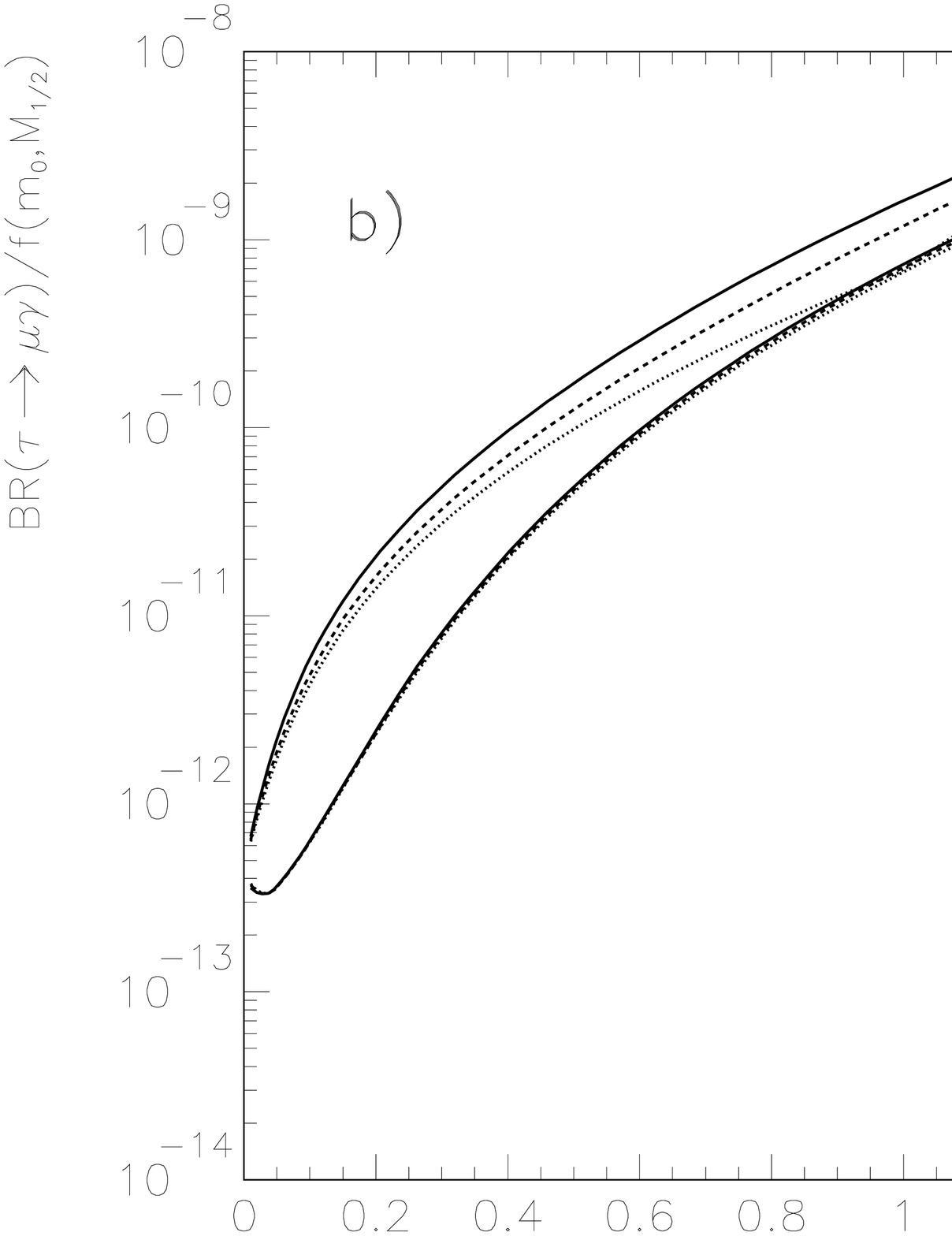}
\caption{\it Extremal values of $BR(\tau\to\mu\gamma)/f(m_0,M_{1/2})$ as 
a function of $|z|$ for the choice of the parameters described in Table 
\ref{table1}, $\tan\beta =10$ and $A_0=0$. Dotted, dashed and solid lines 
correspond to $\arg z=0,\pi/4,\pi/2$, respectively. \label{fitmg2}}
\end{figure}

This behaviour is clearly seen 
in Fig.~\ref{fitmg1}, which shows $\btmg$ as a function of $\phi_2$ for 
$|z|^2=1/2$, $\tan\beta=10$ and $A_0=0$, for the two different values of 
the 
$\Un{13}$ element specified in Table~\ref{table1}. As already mentioned, 
the values of 
the branching ratio for other values of $m_0$ and $M_{1/2}$ can be 
obtained by appropriate rescalings that can be read off from 
Fig.~\ref{fiffun}a. The dependence of $BR(\tau\to\mu\gamma)$ on $|z|$ is 
depicted in Fig.~\ref{fitmg2}. For each value of $|z|$, the maximal and 
minimal predictions for this branching ratio are shown. In particular, the 
allowed range for $|z|^2=1/2$ corresponds to a horizontal squeezing of 
Fig.~\ref{fitmg1}. 

At this point, it is worth summarizing the dependence of 
$BR(\tau\to\mu\gamma)$ on the model parameters. The dependence on $|z|$
gives a factor that varies by $10^4$ for $|z|$ changing between 0.01 and
1.4. As can be seen in Fig.~\ref{fiffun}a, for $m_0$ and $M_{1/2}$ varying 
in the range 100~GeV$ \to $1 TeV, the decay rate can change by another a 
factor of roughly $10^4$. Furthermore, the dependence of (\ref{brdef}) on 
$\tan^2\beta$ brings in also a factor of order a few to $\sim 10^3$. 
One should 
also remember the quadratic dependence of $BR(\tau\to\mu\gamma)$ on $M_3$.
The fact that $\phi_2$ is inaccessible to experiment introduces only a small 
additional uncertainty of the order of factor 4.

Collecting all this information, we see that, under the present
assumptions and with the anticipated experimental sensitivity down to
$\btmg\sim 10^{-9}$, this decay can in the scenario (\ref{omegalep}) be
experimentally accessible for a large range of parameters. The question
which part of this parameter space can be consistent with the present
experimental bound $BR(\mu\to e\gamma)<1.2\times10^{-11}$ is considered
below. It is however worth noting that, assuming that the soft masses and
$\tan\beta$ can be determined by other measurements with sufficient
precision, the measurement of this decay would, owing to the strong
dependence on $|z|$: $BR(\tau\to\mu\gamma)\propto |z|^4$, provide a
relatively narrow range of $|z|$ for fixed masses of the heavy singlet
neutrinos.

\begin{figure}
\includegraphics*[height=8cm]{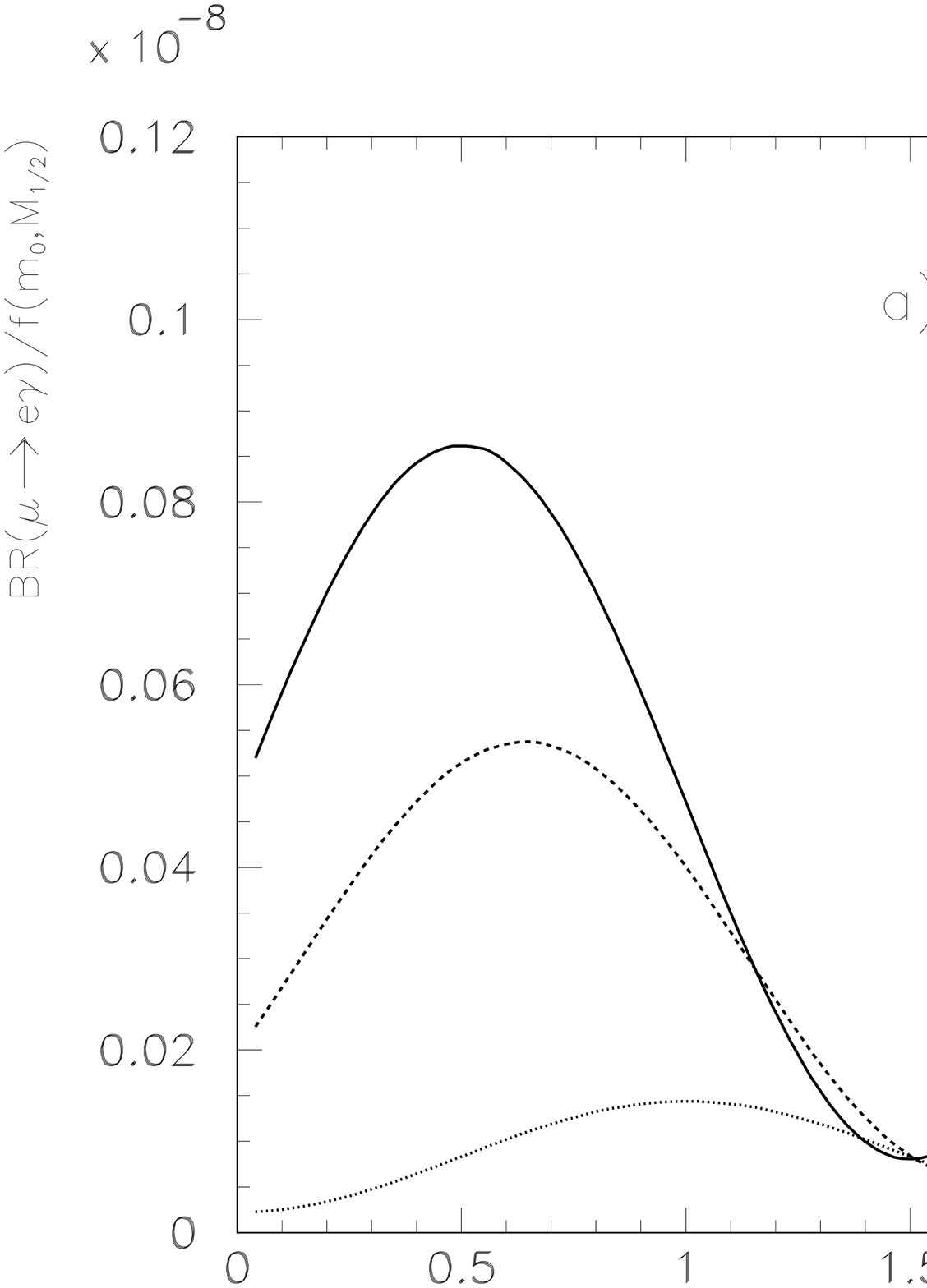}
\hspace{1cm}
\includegraphics*[height=8cm]{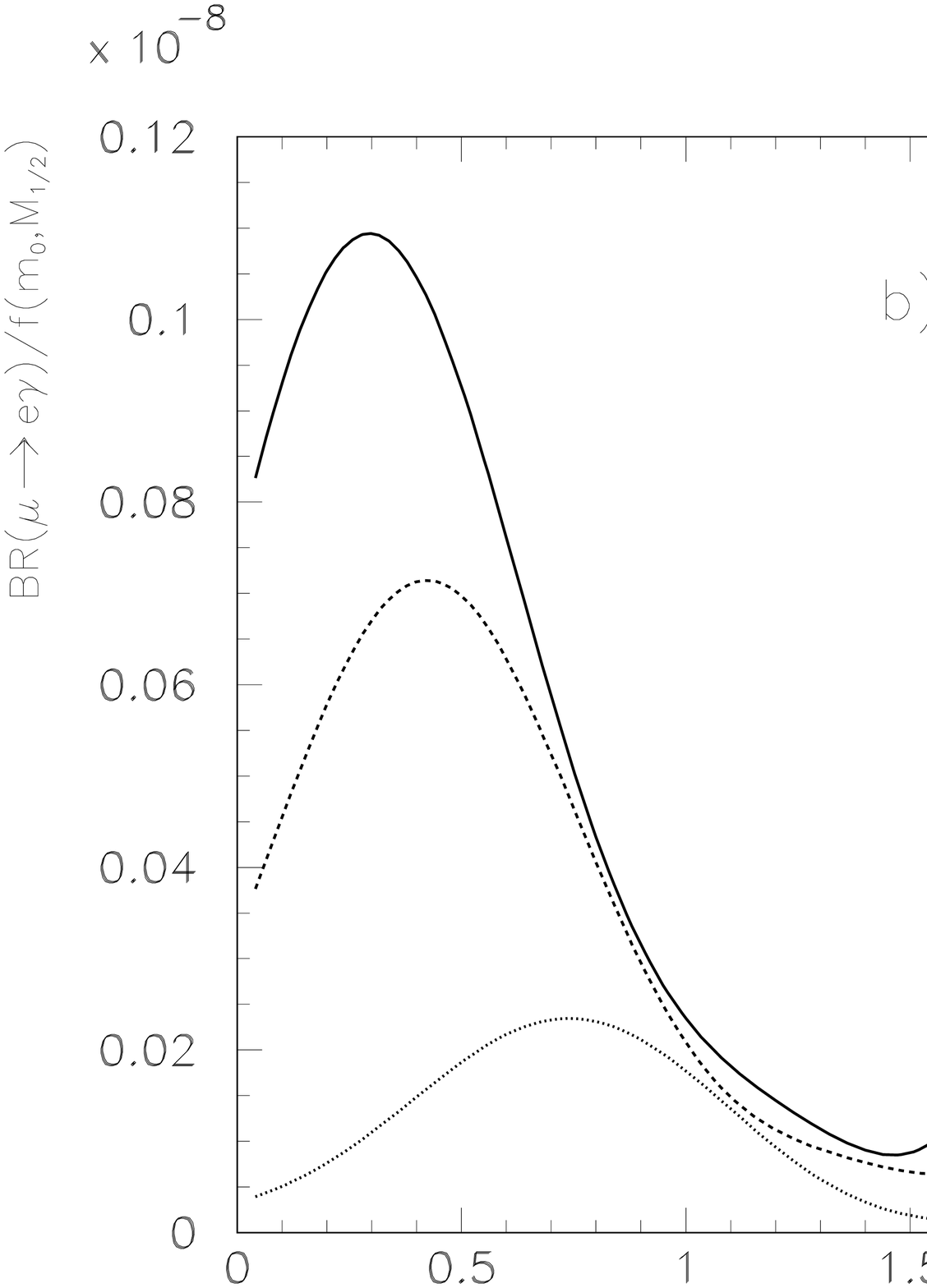}
\caption{\it $BR(\mu\to e\gamma)/f(m_0,M_{1/2})$ as a function of $\phi_2$ 
for the choice of the remaining phases described in Table \ref{table1}, 
$|z|=1/\sqrt2$, $\tan\beta =10$ and $A_0=0$. Dotted, dashed and solid lines 
correspond to $\arg z=0$, $\pi/4$ and $\pi/2$, respectively. \label{fimeg1}}
\end{figure}

The quantities $\msl{21}$ relevant for $\mu\to e\gamma$ and $\msl{31}$ 
relevant for $\tau\to e\gamma$ can be approximated by:
\begin{eqnarray}
\msl{A1}\approx\kappa  
\frac{m_{\nu_3}M_3\Delta t}{\langle H\rangle^2}
\left[R \Un{A3}\Un{12*}\left(Sz\sqrt{1-z^2}^\ast- z^\ast\sqrt{1-z^2}\right) 
\right.\nonumber\\
\left.+ R^2\Un{A2}\Un{12\ast}\left(S|z|^2+|1-z^2|\right) + 
\Un{A3}\Un{13\ast}\left(|z|^2+S|1-z^2|\right)\right], \label{q21}
\end{eqnarray}
where we have dropped the terms suppressed by both $\Un{13}$ and $R$. For 
the sake of qualitative discussion, the third term can always be neglected 
(for $\Un{13}\neq0$, this 
is a good approximation provided $|1-z^2|$ is not too close to zero). For 
$\bmeg$, the factors depending on the neutrino masses and mixing in 
the first two terms in (\ref{q21}) can be estimated as:                    
\beq
R|\Un{23}\Un{12}|\sim 0.16 \qquad R^2|\Un{22}\Un{12}|\sim 0.05,\label{estim}
\eeq
whose ratio is $3.1\pm 0.5$. For $S\ll1$ the relative phase between the 
first two terms in (\ref{q21}) is $\phi_2-\arg(z^\ast\sqrt{1-z^2})+\pi$. 
This is clearly seen in Figs.~\ref{fimeg1}a and \ref{fimeg1}b, where we 
show $\bmeg/f(m_0,M_{1/2})$ as a function of the Majorana phase $\phi_2$ 
for $|z|=1/\sqrt2$ and three representative values of $\arg z$. The 
strength
of the $\phi_2$-dependent interference pattern seen in Fig.~\ref{fimeg1} 
varies with the values of $|z|$ and $\arg z$. In particular, it decreases
both for $|z|\rightarrow0$ and $|z|\gg1$, where the second term in 
(\ref{q21}) 
dominates over the first one. Comparison of 
Fig.~\ref{fimeg1}a with \ref{fimeg1}b shows that a non-zero value of 
$\Un{13}$ enhances the interference pattern but does not change 
it qualitatively.

\begin{figure}
\includegraphics*[height=8cm]{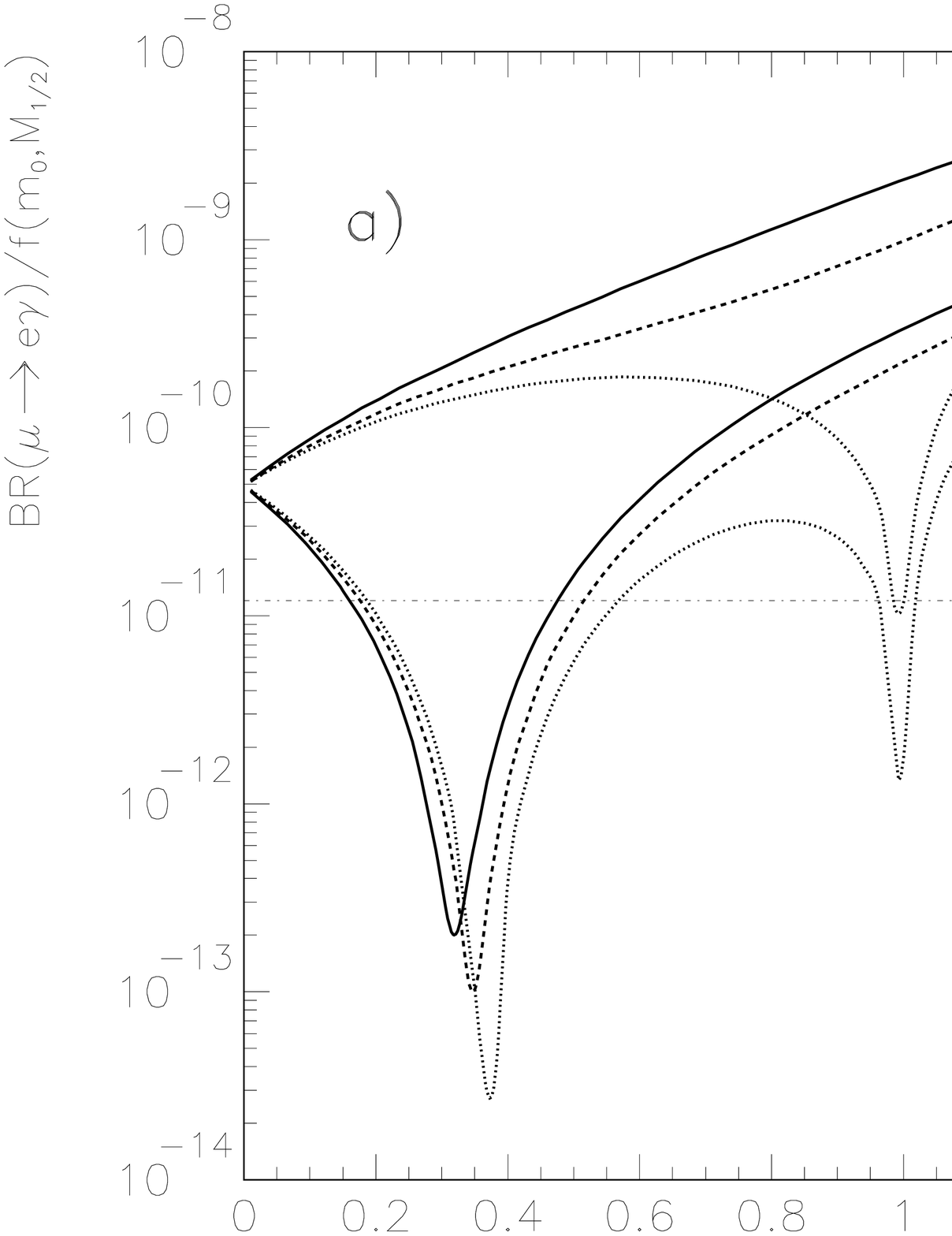}
\hspace{1cm}
\includegraphics*[height=8cm]{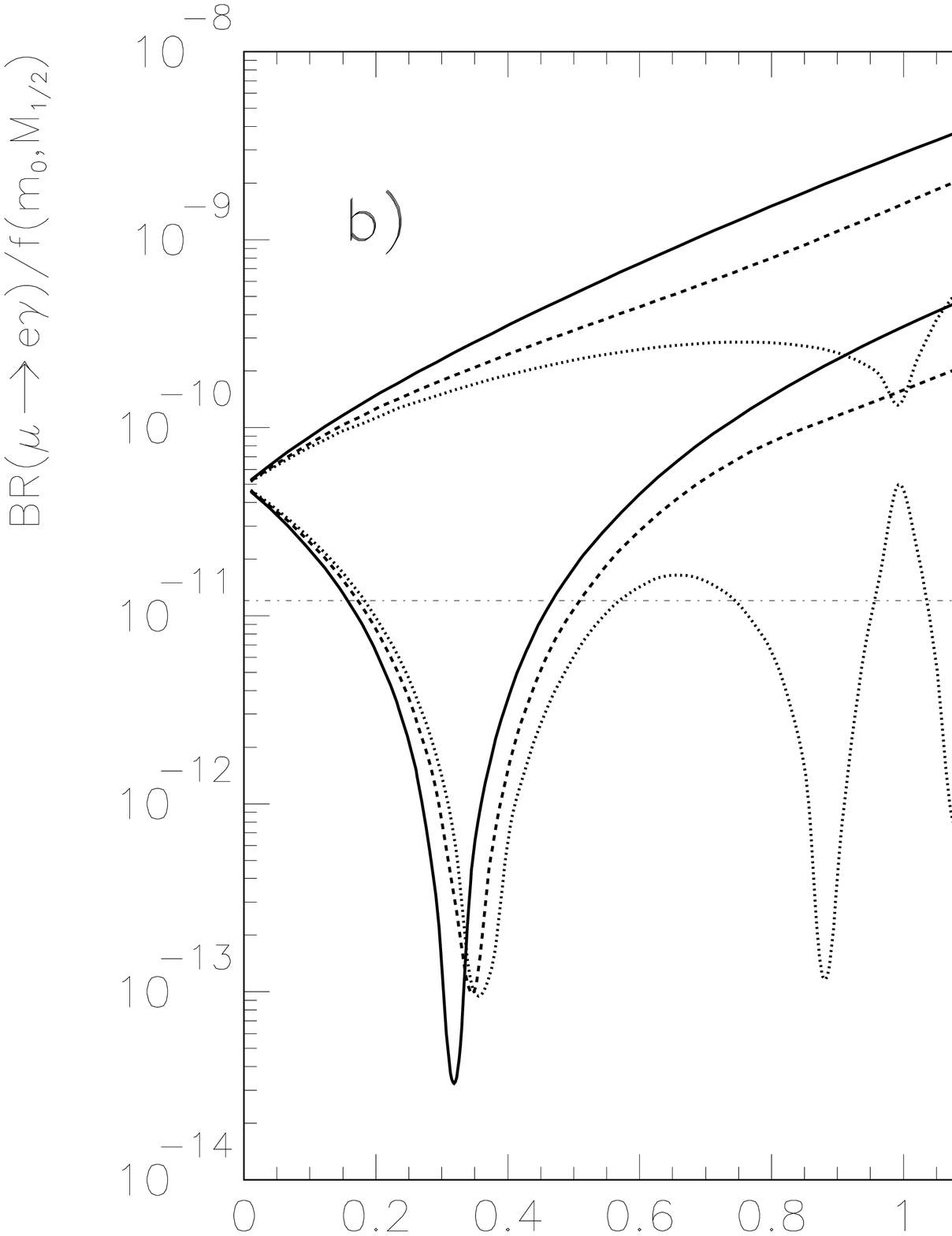}
\caption{\it Extremal values of $BR(\mu\to e\gamma)/f(m_0,M_{1/2})$ as a 
function of $|z|$ for the choice of the parameters described in Table 
\ref{table1}, $\tan\beta =10$ and $A_0=0$. Dotted, dashed and solid lines 
correspond to $\arg z=0$, $\pi/4$ and $\pi/2$, respectively. \label{fimeg2}}
\end{figure}

The dependence of $\bmeg$ on the other parameters is similar to that of
$\btmg$.  In Fig.~\ref{fimeg2} we show $\bmeg$ as a function of $|z|$ for
the same values of the other parameters as we used in Fig.~\ref{fitmg2}
for $\btmg$. The comparison of Figs.~\ref{fiffun} and \ref{fimeg2} shows
that, except for two special points at $|z|\sim0.3$ and $|z|\sim1$, the 
branching ratio of the decay $\mu\to e\gamma$ for 
$f(m_0,M_{1/2})\simgt1$ and $\tan\beta\simgt 10$ is above the experimental 
upper bound $1.2\times 10^{-11}$. 
The minimum of $\bmeg$ at $|z|\sim0.3$ is due to the destructive interference 
of the first two terms in (\ref{q21}): for $S\ll1$ they are of equal 
magnitude just for $|z|\sim0.3$, and for any value of $\arg z$ the 
Majorana 
phase $\phi_2$ can be chosen so that these two terms approximately cancel each 
other. The  minima for $|z|\sim1$ are instead caused by the simultaneous 
vanishing of the two first terms in the limit $S=0$. 
Therefore they appear only for $\arg z\approx0$ and $\pi$. For 
$\mathbf{U}_\nu^{13}=0$ the non-zero value of $\bmeg$ at the single 
minimum 
seen in Fig.~\ref{fimeg2}a is due to $S\neq0$ in the second term in 
(\ref{q21}); for $\mathbf{U}_\nu^{13}\neq0$ (Fig.~\ref{fimeg2}b) there are 
two minima whose position is determined by the cancellation of the first
and third terms in (\ref{q21}), which holds for 
$z\approx\pm(1-{1\over2}\alpha\exp(2i(\delta-\phi_2))$, where 
$\alpha=|\mathbf{U}_\nu^{13\ast}/R\mathbf{U}_\nu^{12\ast}|\approx1/5$,
i.e. for $|z|\approx1-{1\over2}\alpha\cos(2(\delta-\phi_2))$. The minima of 
$\bmeg$ occur at the values of $z$ corresponding to the texture zeroes
(\ref{y22tz}) and (\ref{y32tz}). This is obvious for $\ynu{3A}=0$, since
$BR(l_A\to l_B\gamma)\propto|\ynu{3A\ast}\ynu{3B}|^2$, but it is a
non-trivial result for $\ynu{22}=0$, which follows from the structure of
the masses and mixings of the light neutrinos.

As follows from Fig.~\ref{fiffun}, for $\bmeg$ to be consistent with the
experimental bound in a larger range of $|z|$ values, the gaugino mass
$M_{1/2}$ at the high scale should be bigger than 500 GeV and/or $M_3$
significantly smaller than $5\times10^{14}$ GeV. On the other hand,
$M_{1/2}\simgt500$ GeV corresponds to third-generation masses above 1~TeV
at the electroweak scale, which begins to conflict with the naturalness
requirement. This conflict becomes even sharper for higher values of
$\tan\beta$. This discussion suggests that, if $\OO{}$ is of the
form (\ref{omegalep}), the rate of $\mu\to e\gamma$ decay should be close
to the present experimental bound.

\begin{figure}
\includegraphics*[height=8cm]{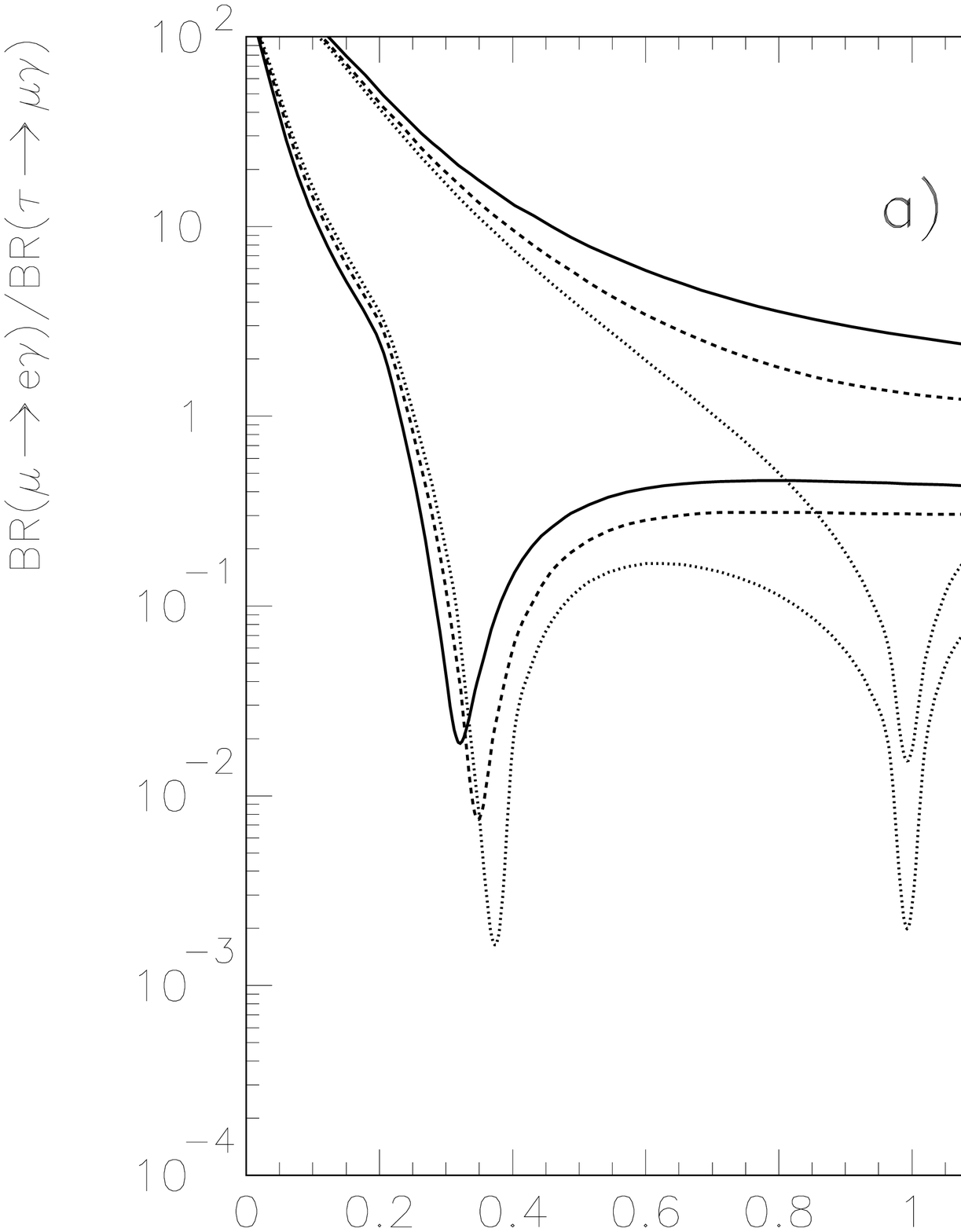}
\hspace{1cm}
\includegraphics*[height=8cm]{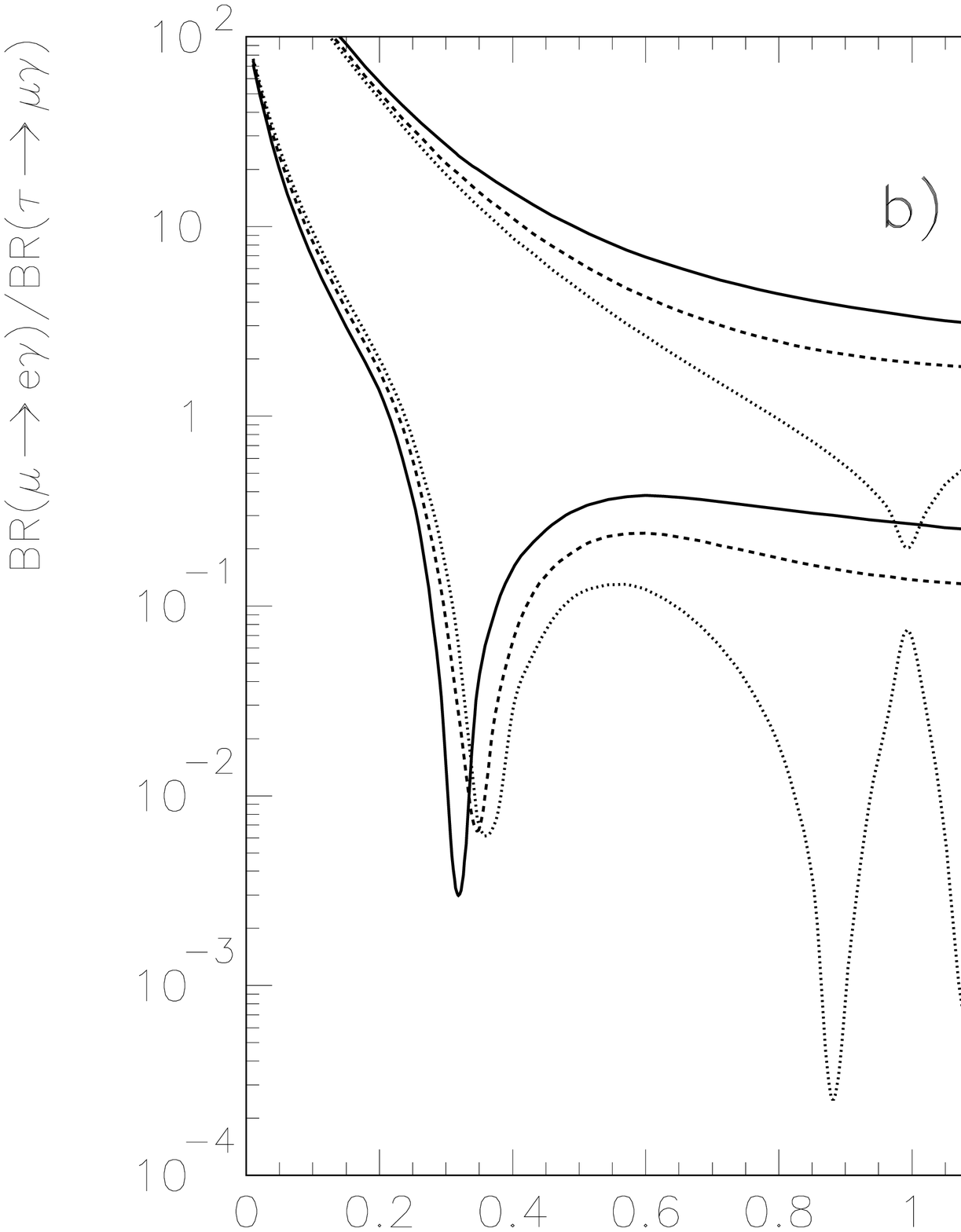}
\caption{\it Extremal values of $BR(\mu\to e\gamma)/BR(\tau\to \mu\gamma)$ 
as a function of $|z|$ for the choice of the parameters described in Table 
\ref{table1}. Dotted, dashed and solid lines correspond to $\arg z=0$,
$\pi/4$ and $\pi/2$, respectively. \label{fimeg3}}
\end{figure}

At this point, we can also address the question whether the present
experimental bound $\bmeg<1.2\times 10^{-11}$ allows for $\btmg\geq
10^{-9}$. This can best be discussed by 
studying the predictions for the ratio $\bmeg/\btmg$, which depends only 
on $z$ and $\phi_2$, as the dependence on the soft supersymmetry-breaking 
parameters, $\tan\beta$ and $M_3$ cancel out. Furthermore,
since $BR(\tau\to\mu\gamma)$ is a monotonic function of $|z|$ and
can be changed by a factor of 10 at most by changing the Majorana phase
$\phi_2$, the minima of the ratio $\bmeg/\btmg$ follow the minima of
$\bmeg$. Maximal and minimal possible values of 
$BR(\mu\to e\gamma)/BR(\tau\to\mu\gamma)$ obtained by varying $\phi_2$ 
in the range $(0,2\pi)$ are shown in Fig.~\ref{fimeg3} as a function of 
$|z|$ for three different values of $\arg z$.

As seen in Fig.~\ref{fimeg3}, generically $BR(\mu\to e\gamma)/
BR(\tau\to\mu\gamma)\sim0.1$ to $1$, that is 
$BR(\tau\to\mu\gamma) \lappeq 10^{-10}$ for $BR(\mu\to e\gamma) \lappeq 
10^{-11}$. However, there are exceptions to this rule in a few isolated 
regions corresponding to the minima of $\bmeg$ seen in Fig.~\ref{fimeg2}. 
This means that observation of $\tau\to\mu\gamma$ at a rate $\simgt 10^9$ 
in future
experiments would be a strong constraint for top-down models of neutrino 
masses and mixings.

\begin{figure}
\includegraphics*[height=8cm]{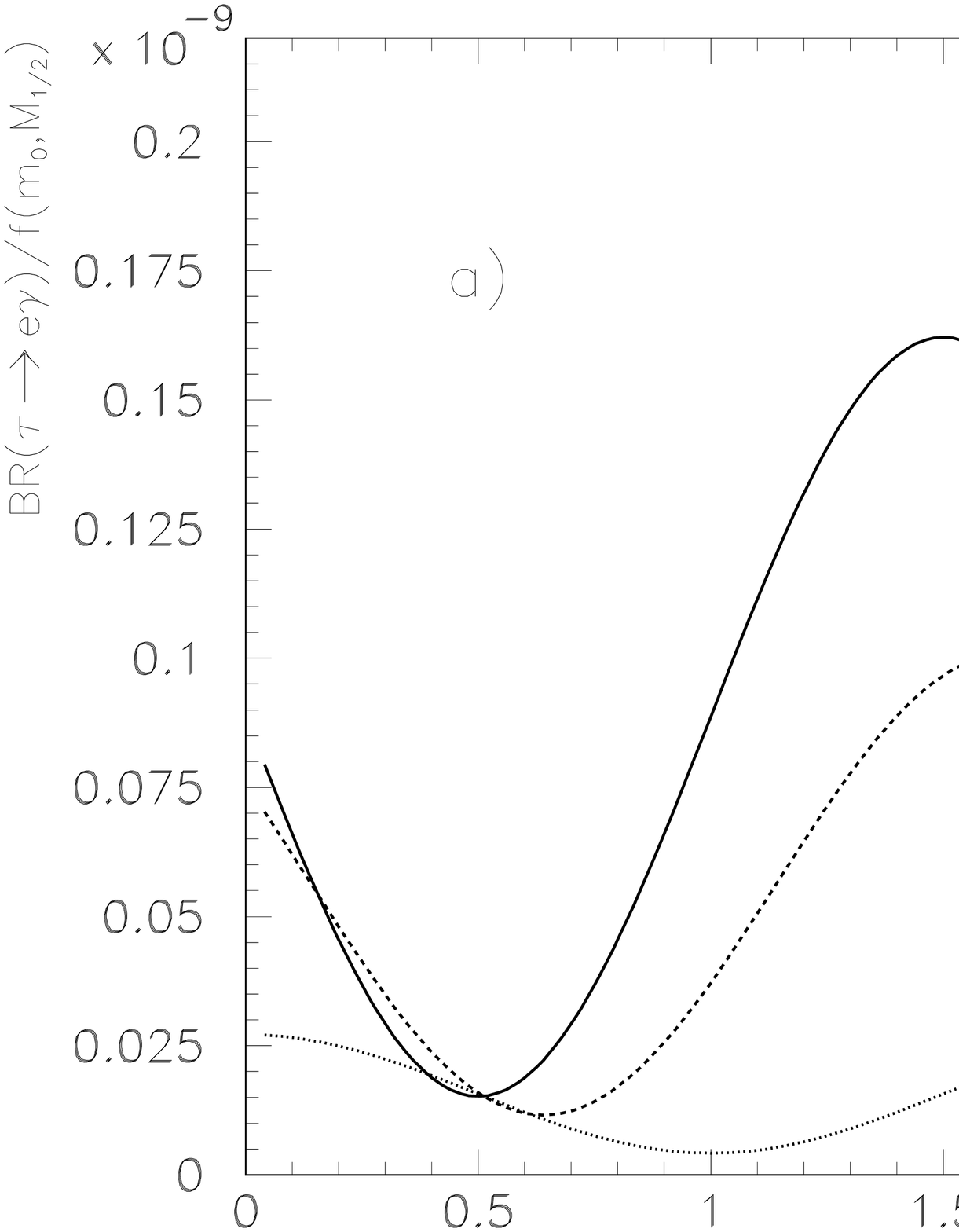}
\hspace{1cm}
\includegraphics*[height=8cm]{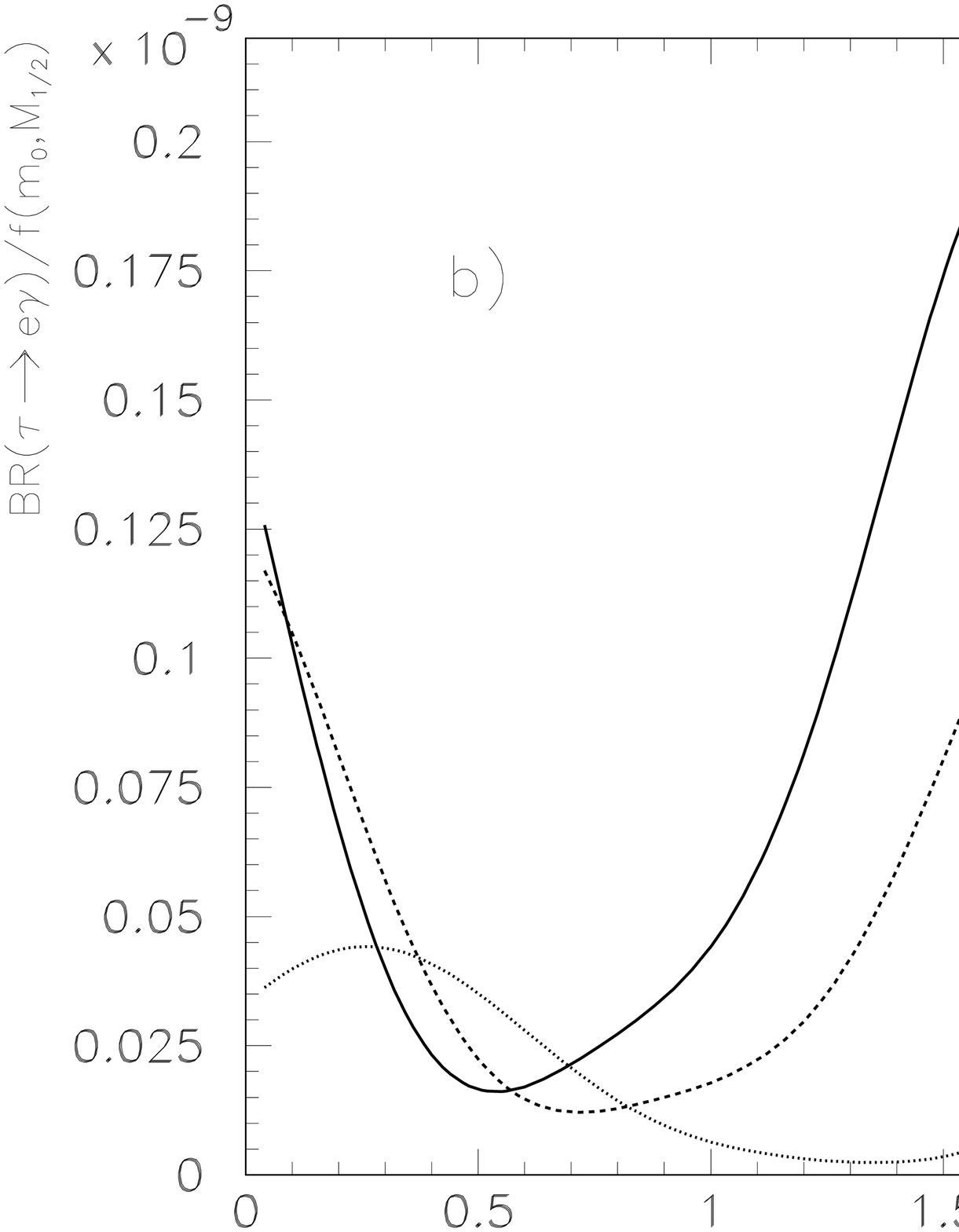}
\caption{\it $BR(\tau\to e\gamma)$ as a function of $\phi_2$ for the 
choice of 
the remaining phases described in Table \ref{table1}, $|z|=1/\sqrt2$, 
$\tan\beta=10$ and $A_0=0$. Dotted, dashed and solid lines correspond to 
$\arg z=0$, $\pi/4$ and $\pi/2$, respectively. \label{fiteg1}}
\end{figure}

\begin{figure}
\includegraphics*[height=8cm]{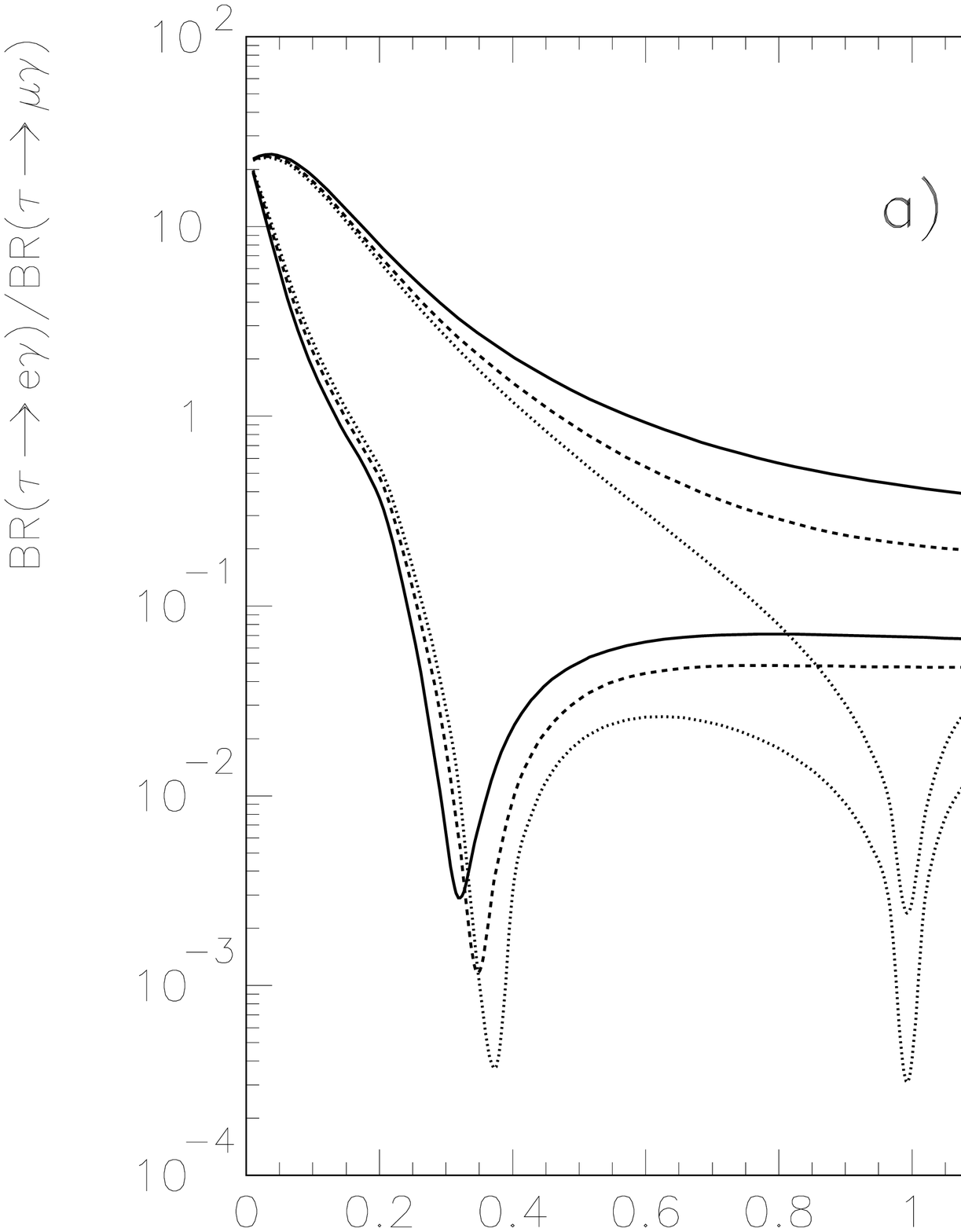}
\hspace{1cm}
\includegraphics*[height=8cm]{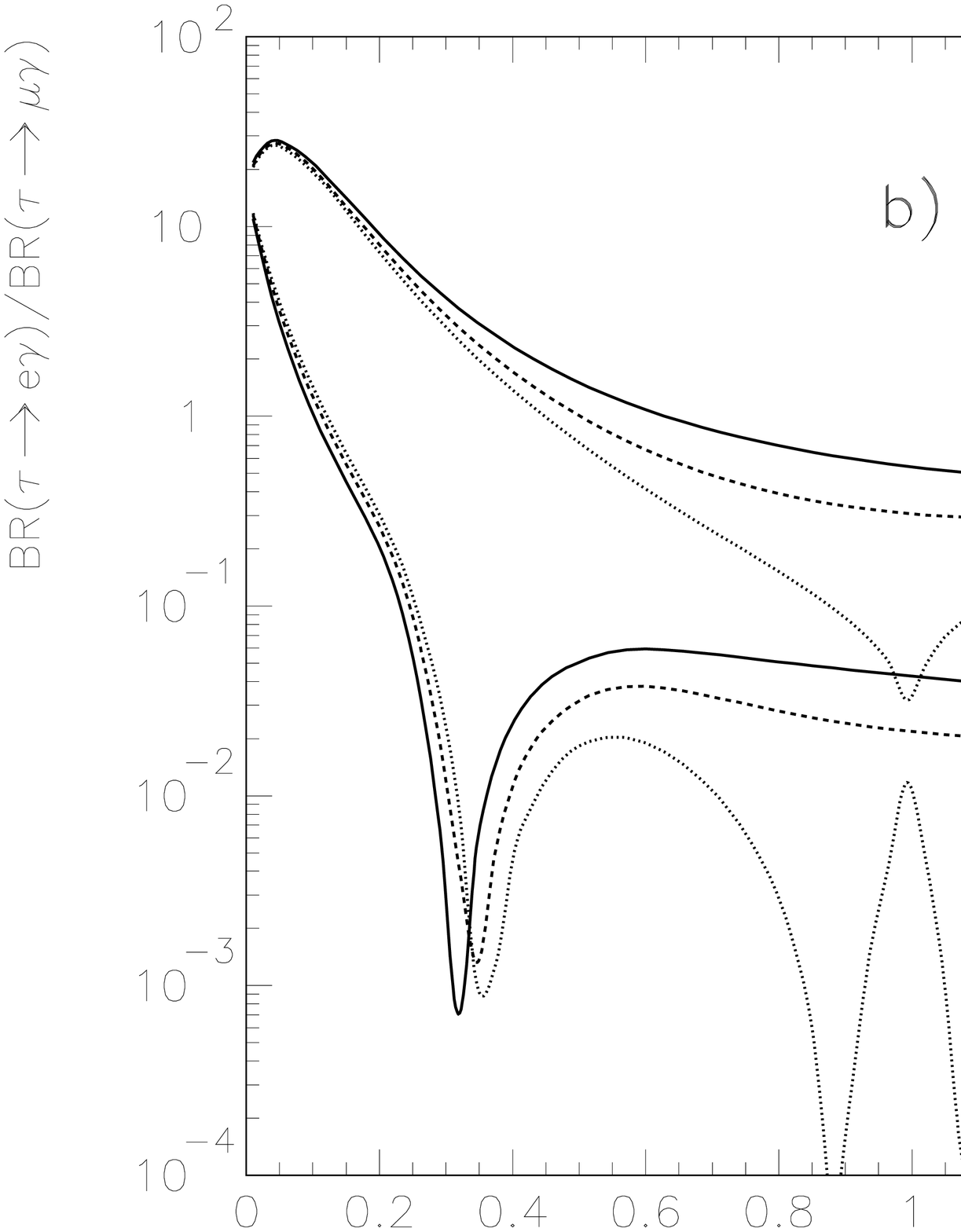}
\caption{\it The extremal values of $BR(\tau\to e\gamma)/BR(\tau\to 
\mu\gamma)$ 
as a function of $|z|$ for the choice of the parameters described in Table 
\ref{table1}, $\tan\beta =10$ and $A_0=0$. Dotted, dashed and solid lines 
correspond to $\arg z=0$, $\pi/4$ and $\pi/2$, respectively. \label{fiteg2}}
\end{figure}

Finally, we briefly consider the decay $\tau\to e\gamma$. The dependences
of the predictions for $BR(\tau\to e\gamma)$ on the soft mass parameters
$z$ and $\phi_2$ are similar to those of $BR(\mu\to e\gamma)$. However,
since the relative phase between $\Un{22}$ and $\Un{32}$ approximately
equals $\pi$, for the value of $\phi_2$ for which $BR(\mu\to e\gamma)$ is
minimized $BR(\tau\to e\gamma)$ is maximal and vice versa, as shown in
Fig.~\ref{fiteg1}. The dependence of $BR(\tau\to
e\gamma)/BR(\tau\to\mu\gamma)$ is depicted in Fig.~\ref{fiteg2}.

\begin{figure}
\hspace{1cm}
\includegraphics*[height=8cm]{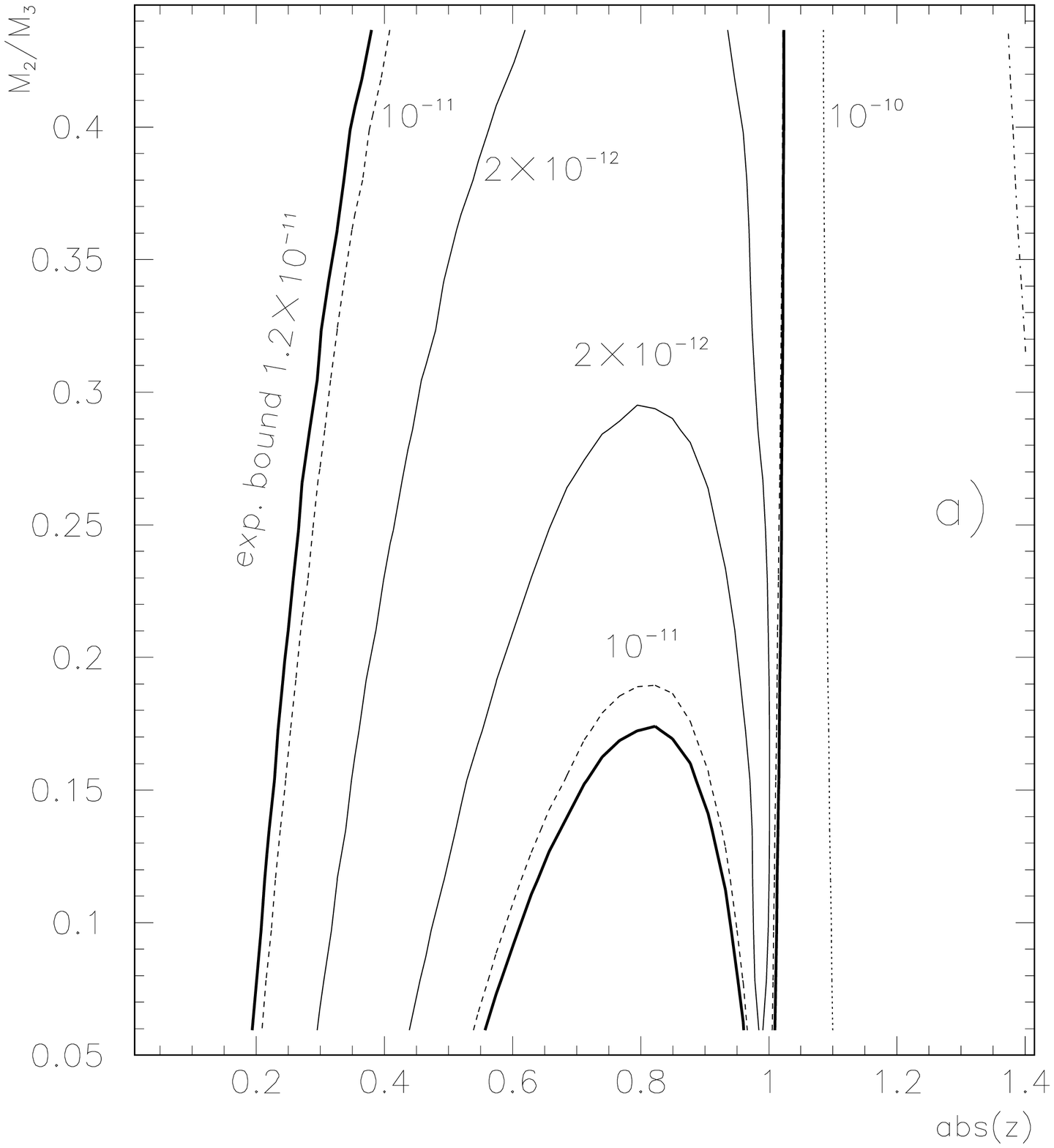}
\hspace{1cm}
\includegraphics*[height=8cm]{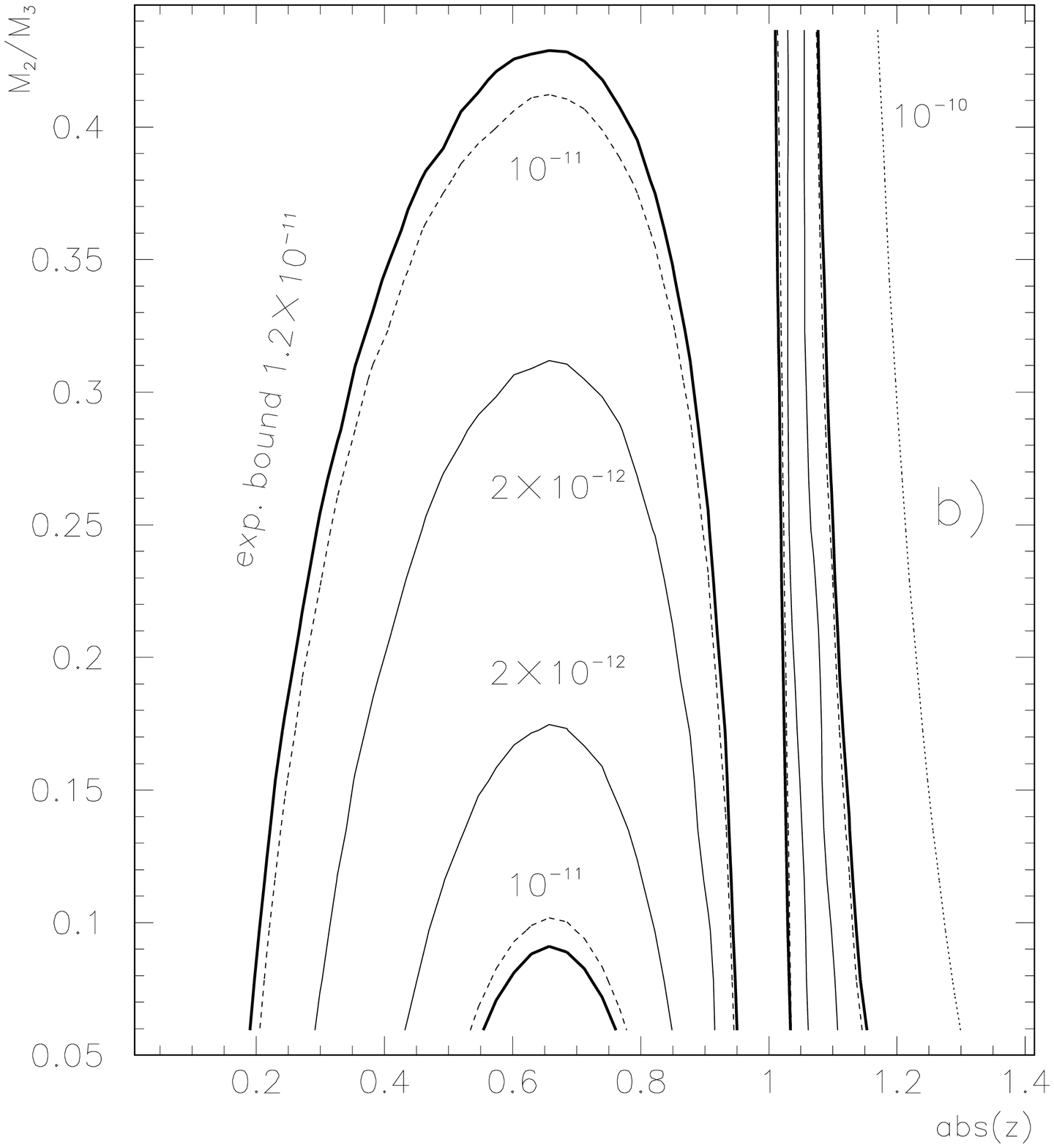}
\caption{\it Contour plot of $BR(\mu\to e\gamma)/f(m_0,M_{1/2})$ minimized 
with 
respect to the variation of $\phi_2$ as a function of $|z|$ and $M_2/M_3$ for 
the choice of the parameters described in Table \ref{table1}, $\tan\beta =10$, 
$A_0=0$ and $\arg z=0$. 
\label{fismeg}}
\end{figure}

We have presented predictions obtained under the supplementary assumption
(\ref{assumpt2}). However, for $M_2$ close enough to $M_3$ but still
reasonably smaller, the first condition (\ref{assumpt2}) is no longer
satisfied. It is therefore worthwhile to check the behaviour of the
predictions for the decays rates when $M_2$ approaches $M_3$. In
particular, one may wonder whether the deep minima in $\bmeg$ at $|z|\sim
0.3$ and $|z|\sim 1$ are filled in by additional contributions to
(\ref{RGEsol1}) coming from the product $(\ynu{2A})^\ast\ynu{2B}$, which 
may now be non-negligible. This is shown in Fig.~\ref{fismeg}. 
The two minima persist even for relatively large values
of the mass ratio, $M_2/M_3\simlt0.4$ for $\mathbf{U}_\nu^{13}=0$ and
$M_2/M_3\simlt0.25$ for $|\mathbf{U}_\nu^{13}|=0.1$, but the first minimum 
shifts to slightly higher values of $|z|$ for larger values of $M_2/M_3$.

In this Section we have discussed the results for $BR(l_A\to l_B\gamma)$ 
obtained under the hypothesis that the lightest right-chiral neutrino 
decouples from the seesaw mechanism. To a very good approximation, these 
results do  not depend on the mass $M_1$. The particular value
$M_1=2\times10^{13}$ GeV, is compatible with inflation driven by the 
lightest sneutrino. As analyzed in \cite{ELRAYA2}, succesful nonthermal 
leptogenesis with low reheating temperature can then occur.  This is because 
for the zeroes in (\ref{omegalep}) representing small complex numbers it 
is easy to obtain values of the CP asymmetry parameter $\epsilon_1$ saturating 
the Davidson-Ibarra upper bound \cite{DIbound}\footnote{However, for moderately 
hierarchical heavy neutrino masses it is possible to obtain larger values of the 
CP asymmetry than the abovementioned bound \cite{HLNPS}.}. 

For any other value, $M_1\neq 2\times10^{13}$ GeV but still with the
pattern (\ref{omegalep}), some other field must be responsible for inflation.
Disregarding the cosmological gravitino problem, one may then contemplate 
the possibility of conventional thermal leptogenesis. It turns out that, with
zeroes in (\ref{omegalep}) representing small complex numbers, 
leptogenesis can be 
realized for $T_{RH}\simgt10^9$ GeV~\cite{BUDIPL,GINORARIST}. In both cases, the 
final 
lepton number asymmetry does not depend on the parameter $z$ of the $\OO{}$ 
matrix. The possibility of lowering the reheating temperature necessary for 
the thermal leptogenesis in a scenario with degenerate neutrinos $N_1$ 
(decoupled) and 
$N_2$ (not decoupled) remains an open question for the pattern 
(\ref{omegalep}).


\subsection{LFV decays and the decoupling of $N_2$ or $N_3$.}
\label{seclfv2}

The predictions for the LFV decays with the decoupling of $N_2$ [pattern 
(\ref{omegatwo})] or $N_3$ [pattern (\ref{omegadec})] are easy to discuss 
if one remembers that $\msl{AB}$ in eq. (\ref{RGEsol1}) is dominated by 
the contribution of the product $(\ynu{3A})^\ast\ynu{3B}$ or 
$(\ynu{2A})^\ast\ynu{2B}$, respectively. This means that 
in the case (\ref{omegatwo})
the predictions for the decay rates are the same as for pattern
(\ref{omegalep}), for the same values of $\ynu{3A}$ and if 
in both cases the subleading contributions to $\msl{AB}$ are 
negligible~\footnote{Predictions in the two patterns are the same 
including 
the subleading effects if the $\ynu{1A}$ in the case (\ref{omegatwo}) are
numerically equal to the $\ynu{2A}$ in the case (\ref{omegalep}).}. For 
the 
decoupling of $N_3$, the predictions are the same as for the decoupling of 
$N_1$ if we change $\ynu{3A}\to\ynu{2A}$ and keep the numerical values
unchanged. At this point, it is worth recalling equation 
(\ref{YukawaOmega}):
\bea
\ynu{3A} &=& iM_3^{1/2} \left( \OO{31}c_1\Un{A1*}+ \OO{32}c_2\Un{A2*}
+ \OO{33}c_3\Un{A3*}\right) \nonumber\\
\ynu{2A} &=& iM_2^{1/2} \left( \OO{21}c_1\Un{A1*}+ \OO{22}c_2\Un{A2*}
+ \OO{23}c_3\Un{A3*}\right)
\label{starsp}
\eea
We see that the predictions following from decoupling of $N_1$ and $N_2$ 
are, to a good approximation, identical for the same
values of $M_3$. In particular,
the mass of the decoupled neutrino can be kept at $2\times10^{13}$~GeV (for 
sneutrino inflation)
\footnote{With the exception that, in the case (\ref{omegatwo}), 
subleading effects 
may become important only at the expense of lowering $M_3$, i.e., when the 
LFV decay rates are suppressed.}. Decoupling of $N_3$ leads to the same
LFV decay rates as decoupling of $N_1$ for $M_2$ replaced by a
numerically equal $M_3$. Moreover, with $M_3=2\times 10^{13}\gev$
the dominant contribution to $\ynu{2A}$ in 
(\ref{starsp}) comes from $M_2<M_3$. So the $BR(l_A\to l_B\gamma)$
discussed in Section \ref{seclfv} are rescaled by the ratio $M_2^2/M_3^2$.
Since, for (\ref{omegalep}), $M_3\gg 2\times 10^{13}\gev$ and, for
(\ref{omegadec}), $M_2\ll M_3=2\times 10^{13}\gev$, the predictions for
the decay rates are strongly suppressed in the latter case. 
We conclude that, if the heaviest 
right-chiral neutrino is the inflaton, the decays $l_A\to l_B\gamma$
are unlikely to be observed experimentally.

\begin{figure}
\includegraphics*[height=8cm]{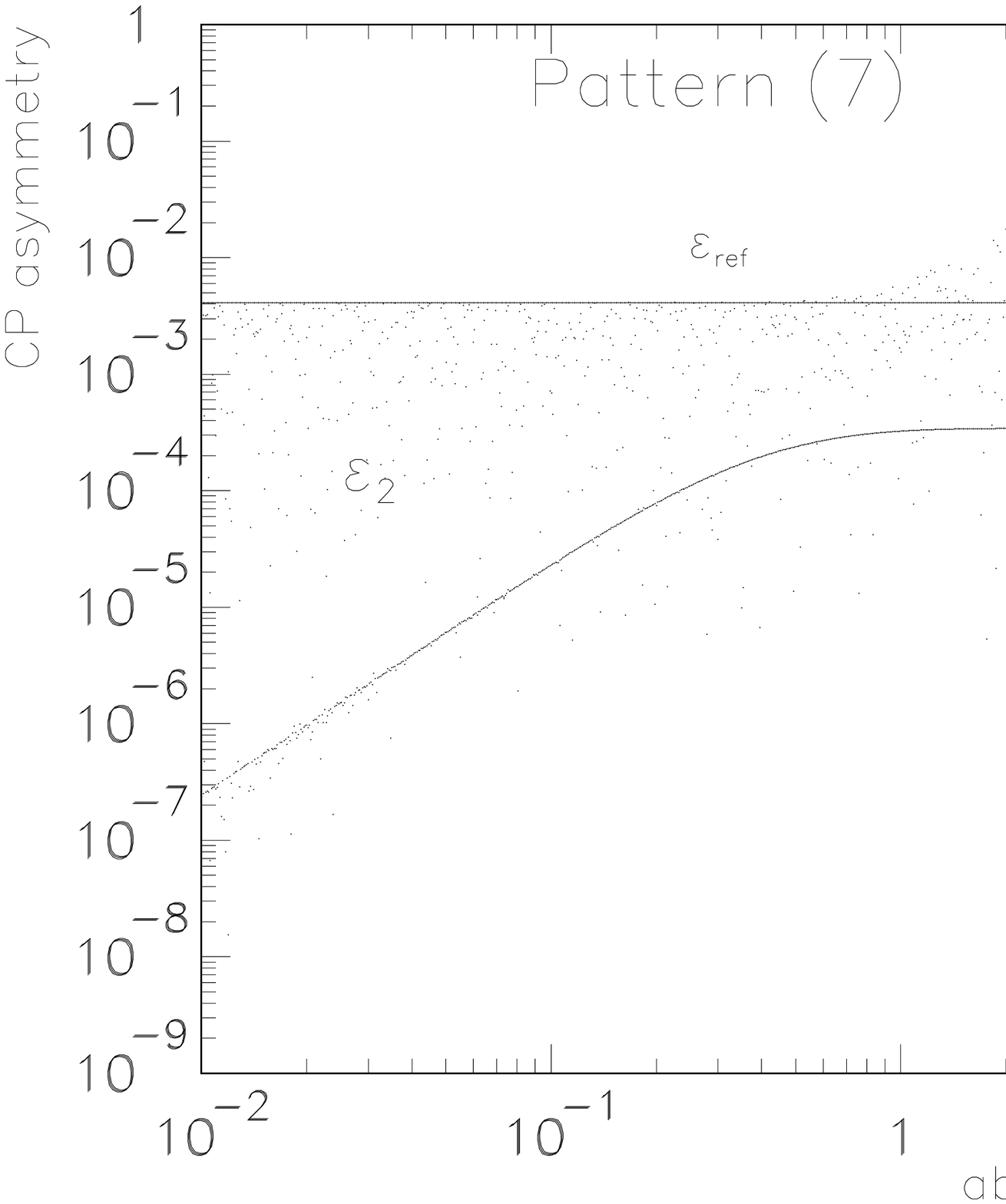}
\hspace{1cm}
\includegraphics*[height=8cm]{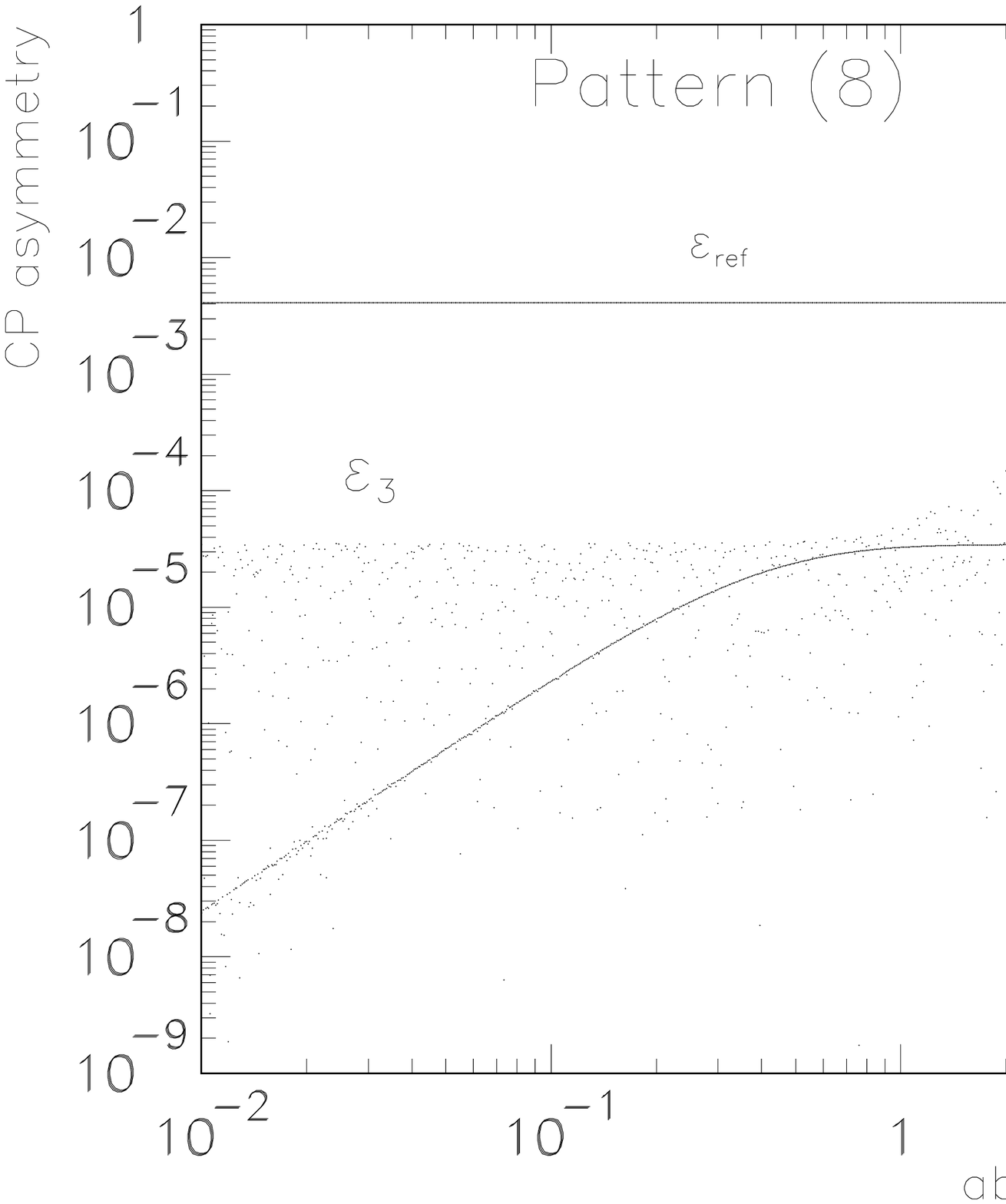}
\caption{\it The CP asymmetries generated in the decays of the 
sneutrino-inflaton for 
the pattern (\ref{omegatwo}) (left panel) and (\ref{omegadec}) (right 
panel). Details of 
the scanning procedure are given in the text.
\label{fistab}}
\end{figure}

Let us now consider the possible types of leptogenesis with $N_2$ or $N_3$ 
decoupled. In nonthermal leptogenesis following sneutrino-driven
inflation~\cite{ELRAYA2}, the final lepton asymmetry is generated 
directly in the decay of the vacuum condensate of the sneutrino-inflaton 
field
$\tilde N_2$ or $\tilde N_3$. Since the final lepton-number-to-entropy 
ratio
$Y_L$ generated in this way is given by:
\beq
Y_L = {3\over4}\epsilon_A\frac{T_{RH}}{M_\mathrm{infl}}~,
\eeq 
a low reheating temperature $T_{RH}\simlt10^7$ GeV and 
$M_\mathrm{infl}= 2\times10^{13}$~GeV require the CP asymmetry 
parameters $\epsilon_2$ or $\epsilon_3$, respectively, to assume 
values of the order of $\epsilon_\mathrm{ref}/10$, where 
$\epsilon_\mathrm{ref}=\frac{3M_\mathrm{infl}m_{\nu_3}}{8\pi\langle H\rangle^2}\sim 4\times 10^{-3}$ 
(for $\tilde{N}_1$-driven inflation $\epsilon_\mathrm{ref}$ is the same
as the Davidson-Ibarra bound on $\epsilon_1$ \cite{DIbound}). 
In Table \ref{table2} we display 
the one-loop contributions to $\epsilon_A/\epsilon_\mathrm{ref}\eta_{AE}$
from the individual virtual 
states $N_E$ (and $\tilde N_E$) \cite{COROVI}, where
\beq
\eta_{AE}=\left|\frac{\sum_D m_{\nu_D}^2 \mathsf{Im}\left[(\OO{AD}\OO{ED}^*)^2\right]}{m_{\nu_3}\sum_Cm_{\nu_c}|\OO{AD}|^2}\right|\leq \sum_D\frac{m_{\nu_D}}{m_{\nu_3}} |\OO{ED}|^2
\label{etabound}
\eeq
For the pattern (\ref{omegatwo}), only 
the contribution from $N_1$ is suppressed by the small value of $(M_1/M_2)^2$, 
whereas for the pattern (\ref{omegadec}) all the contributions are small. 
This suppression can only be overcome if $\eta_{AE}$, that is some 
$|\OO{ED}|^2$, are large, which, 
recalling the interpretation of $\OO{}$ as the dominance matrix, means some 
fine-tuning in the seesaw mechanism. This can be seen in Figure~\ref{fistab}, 
where we show in the form of a scatter plot the CP asymmetries 
$\epsilon_2$ 
and $\epsilon_3$ generated in the direct decays of the sneutrino-inflaton 
$\tilde{N}_2$ [$\tilde{N}_3$],  relevant for the patterns (\ref{omegatwo}) 
[(\ref{omegadec})], respectively, as functions of the parameter 
$p=\sqrt{1-z^2}$ of the $\OO{}$ matrix. We have chosen $\arg p=\pi/4$ and 
$\arg\OO{B2}=\arg\OO{B3}=\pi/4$ for $B=2,3$ and scanned over $|\OO{B2}|$ and 
$|\OO{B3}|$ in the range $10^{-3}-10^{-1}$ with a uniform distribution 
over the logarithmic scale. It is 
clear from Figures~\ref{fistab}a and \ref{fistab}b
that, whilst the Davidson-Ibarra bound~\cite{DIbound} can be saturated 
for the pattern (\ref{omegatwo}) with $|z|,|p|\simlt1$, for the pattern 
(\ref{omegadec}) this is possible only for $|z|,|p|\simgt3$ and, hence, 
requires fine tuning in the seesaw mechanism.

\begin{table}
\begin{center}
\begin{tabular}{|c|c|c|}
\hline
 & pattern (\ref{omegatwo}) & pattern (\ref{omegadec}) \\
\hline
$E=1$ & $\frac{2M_1^2}{3M_2^2}\left(1+\ln\frac{M_1}{M_2}\right)$ 
& $\frac{2M_1^2}{3M_3^2}\left(1+\ln\frac{M_1}{M_3}\right)$ 
\\
$E=2$ & $---$ & $\frac{2M_2^2}{3M_3^2}\left(1+\ln\frac{M_2}{M_3}\right)$ 
\\
$E=3$ & $1$ & $---$ \\
\hline
\end{tabular}
\end{center}
\caption{\it Estimated maximal contributions to
$|\epsilon_2/\epsilon_\mathrm{ref}\eta_{2E}|$ for the pattern (\ref{omegatwo}) 
and to $|\epsilon_3/\epsilon_\mathrm{ref}\eta_{3E}|$ for the pattern 
(\ref{omegadec}), due to virtual $N_E$ exchange in the loop (see the text
for explanation).} 
\label{table2}
\end{table}

\begin{figure}
\includegraphics*[height=8cm]{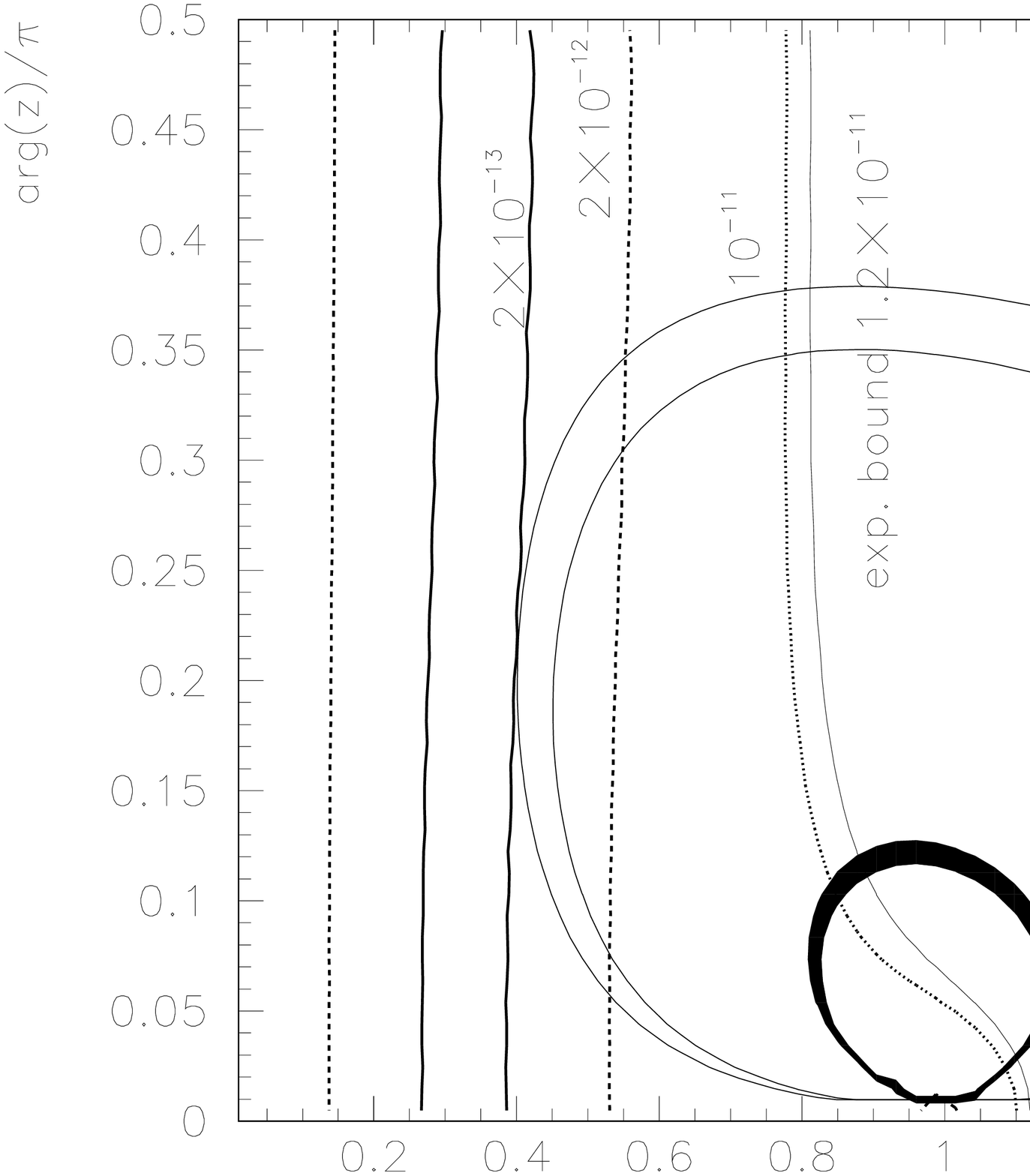}
\hspace{1cm}
\includegraphics*[height=8cm]{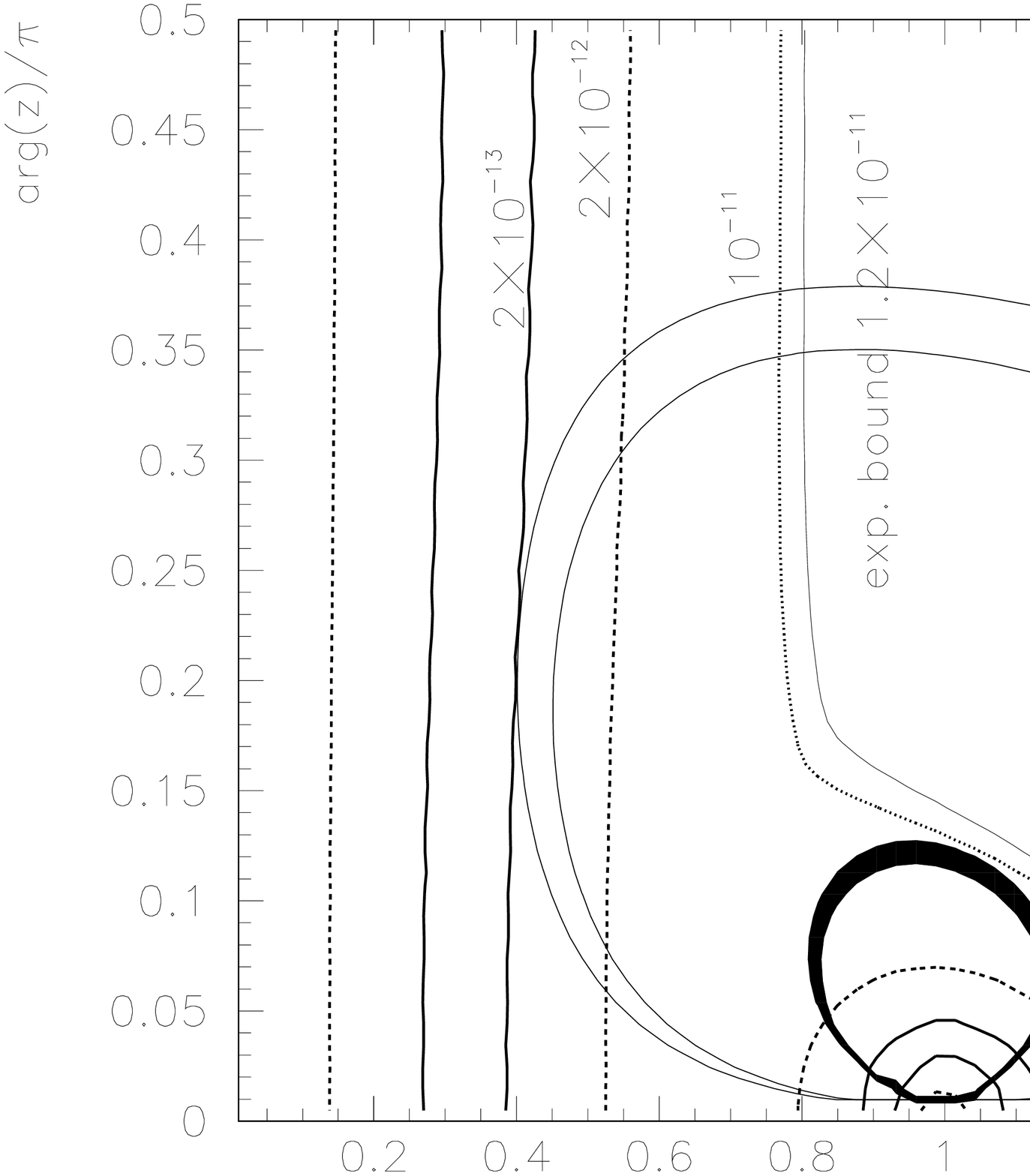}
\caption{\it Minimal values of $\bmeg/f(m_0,M_{1/2})$ for $\tan\beta=10$
and $A_0=0$. The dark ring is the region of $z$ 
in the complex plane for which leptogenesis is successful for 
$M_1=10^{11}$~GeV. The larger ring bounded by thin solid 
lines show how this region 
changes for $M_1=5\times10^{11}$~GeV.\label{dwaa}}
\end{figure}

Finally, we can abandon the hypothesis of sneutrino-driven inflation and 
consider more conventional thermal leptogenesis with the patterns
(\ref{omegatwo}) and (\ref{omegadec}). The masses of all the three right-chiral
neutrinos, including the decoupled one, can be then arbitrary. If they are
split sufficiently for the final lepton number asymmetry to be generated
entirely in the decays of the lightest right-chiral neutrino, the only
relevant CP asymmetry parameter is $\epsilon_1$, which in both cases
(\ref{omegatwo}) and (\ref{omegadec}), takes the form
\beq
\label{epsdec}
\epsilon_1 \approx -\frac{3M_1}{8\pi\langle H\rangle^2}\frac{\mathsf{Im}(z^2) 
\left( m_{\nu_3}^2-m_{\nu_2}^2\right)}{|z|^2m_{\nu_2}+|1-z^2|m_{\nu_3}}
\eeq
The important wash-out parameter $\tim$ then reads
\beq
\label{timdec}
\tim\approx\left(|z|^2 m_{\nu_2}+|1-z^2| m_{\nu_3} \right) ~.
\eeq
The basic consequences of (\ref{epsdec}) and (\ref{timdec})
were explored in~\cite{CHTU}.\footnote{In \cite{CHTU} only the 
decoupling of the heaviest right-chiral neutrino, i.e., the pattern 
(\ref{omegadec}), was considered explicitly. However, it is clear 
that these results apply to the pattern (\ref{omegatwo}) as well.}
It was found that, because $\tim$ is bounded from below by $m_{\nu_2}$
and not by $m_{\nu_1}$, the  wash-out processes are very efficient and 
successful leptogenesis requires compensation by a large $\epsilon_1$, 
i.e.,
large $M_1$. For the patterns (\ref{omegatwo}) and (\ref{omegadec}), with
$M_1=10^{11}$ GeV and $M_1=5\times10^{11}$ GeV, regions of the plane 
$(|z|,\arg z)$ leading to successful thermal leptogenesis are shown in 
Fig.~\ref{dwaa}~\footnote{Recent advances in the calculation of the lepton 
number asymmetry~\cite{GINORARIST} have been included.}. The two panels of 
Fig.~\ref{dwaa}
correspond to the two cases specified in Table 1. Contours of constant
$\bmeg$ minimized with respect to $\phi_2$ are also shown. As can be seen,
regions of the $(|z|,\arg z)$ plane where the thermal leptogenesis reproduces 
the observed baryon number asymmetry of the Universe and regions where the
experimental bound on $\bmeg$ can be respected overlap with each other. 
However, since $T_{RH}\simgt M_1\simgt10^{11}$ GeV for this mechanism to 
work, the cosmological gravitino problem has to solved in some other way,
perhaps by the gravitino being the lightest supersymmetric 
particle~\cite{FUIBYA} (see, however, \cite{ELOLSASP}).

As has been discussed in \cite{CHTU,FLPASA,PIUN,GOFEJONO,ELRAYA1,TU}, for the 
pattern (\ref{omegadec}) $T_{RH}\simlt10^6$ GeV can be obtained if $N_1$ and 
$N_2$ are tightly degenerate, with $1-M_1/M_2\simlt10^{-5}$. In this case, 
however, since $N_3$ is decoupled and $M_2\simlt10^6$ GeV, branching
ratios $BR(l_i\to l_j\gamma)$ are $\sim10^{-18}$, too small to be
accessible to current experiments.
For the pattern (\ref{omegatwo}), one of the two degenerate neutrinos
would be the decoupled one, and to obtain succesful leptogenesis with 
$T_{RH}\simlt10^6$ GeV one would probably need the degeneracy of all the
three right-chiral neutrinos. In this case the LFV decays would 
again be inaccessible.


\section{Conclusions}
\label{seccon}

We have calculated the branching ratios for the decays
$l_A\to l_B\gamma$ in the MSSM under the assumption that the masses
of both the light and heavy singlet neutrinos are hierarchical, and that
one singlet neutrino (not necessarily the heaviest one) decouples from the
seesaw mechanism. For each of the three possible realizations
of the decoupling hypothesis we have also discussed possible leptogenesis
scenarios.

The predictions for $BR(l_A\to l_B\gamma)$ do not
depend on which neutrino decouples. To a very good approximation they
depend on only one non-decoupled neutrino mass, the heavier one. Apart
from the dependence on the soft mass terms and on $\tan\beta$, parameters
that will hopefully be measured in other experiments, the branching ratios
depend strongly on just one complex parameter $z$, which fixes all the
relevant Yukawa couplings. Consequently, our two assumptions lead to
relatively rigid predictions for the decay rates. For
$m_0,M_{1/2}\simlt 1\,\mathrm{TeV}$ and $M_A\simgt 10^{13}\gev$, the decay
rate for $\mu\to e\gamma$ is predicted to be close to the present
experimental bound $1.2\times 10^{-11}$, namely $\bmeg\simgt 10^{-14}$.
However, for $\bmeg<1.2\times 10^{-11}$, the decay rate for
$\tau\to\mu\gamma$ is generically below the anticipated experimental
sensitivity $10^{-9}$, except for some special values of $|z|$. The
ratios $\bmeg/\btmg$ and $\bteg/\btmg$ depend on $z$ only, with the 
dependences
on the other parameters cancelling out. These ratios would therefore be 
particularly interesting
to compare with experiment. 

\vskip0.3cm
\noindent {\bf Acknowledgments}\\
P.H.Ch. was supported by the 
Polish State Committee for Scientific Research Grant 2~P03B~040~24 for 
2003-2005 and the EC Contract HPRN-CT-2000-00152. He would also like to 
thank the CERN Theory Division for hospitality while writing the paper.
The work of S.P. and K.T. was partially supported by the Polish State
Committee for Scientific Research Grant 2~P03B~129~24 for 2003-2005
and the EC Contract HPRN-CT-2000-00148. 
The work of M.R. was supported by the ESF Grants 5135 and 5935 and by the
EC MC contract MERG-CT-2003-503626.
K.T. would like to thank the Michigan Center of Theoretical Physics
and Theory Division at CERN, where parts of this work were done,
for their hospitality and stimulating atmospheres.

\section*{Appendix - Stability of the $\OO{}$ patterns}

In bottom-up evolution, the Yukawa couplings $\mathbf{Y}_\nu^{1B}$,
$\mathbf{Y}_\nu^{2B}$ and $\mathbf{Y}_\nu^{3B}$ enter into the RGEs for 
the soft slepton masses at the scales $M_1$, $M_2$ and $M_3$, respectively.
We see from (\ref{YukawaOmega}) that they are expressed at these respective 
scales in terms of the matrix $\OO{}$, which is taken to have one of the 
scale-independent forms (\ref{omegalep})-(\ref{omegadec}). However, once
the $\mathbf{Y}_\nu^{1B}$ (say) is fixed by (\ref{YukawaOmega}) in terms 
of $\OO{}$ at the scale $M_1$, their scale dependences are controlled by 
the appropriate RGEs, and at the scale $M_2$ they do not satisfy the 
relation (\ref{YukawaOmega}) with scale-independent $\OO{}$. Thus, fixing
$\mathbf{Y}_\nu^{2B}$ at $M_2$ from (\ref{YukawaOmega}) with a
scale-independent orthogonal matrix $\OO{}$ is not a  fully consistent
procedure, as it distorts the relations between $\mathbf{Y}_\nu^{1B}$ and
$\mathbf{Y}_\nu^{2B}$ which follow from the assumption that, e.g., with 
(\ref{omegalep})  $\mathbf{Y}_\nu^{1B}/M_1\to0$ for $B=2,3$. 

We can get some control over the quality of this approximation in the 
following way. Suppose that all three right-chiral 
neutrinos are integrated out at a scale $Q_0$ chosen in such a way that the 
threshold corrections to the relation (\ref{seesaw}) vanish. It is 
easy to verify that in a one-loop approximation $M_1<Q_0<M_3$. We denote 
by $\mathbf{C}_0$ the value at $Q_0$ of the 
coefficient of the dimension-five operator 
${\cal O}_5=-\mathbf{C}_{AB}(L_AH)(L_BH)/2$ related to $\mathbf{C}(M_Z)$ 
directly by the RGEs of the MSSM derived in~\cite{CHPL}. 
At the scale $Q_0$, we can use (\ref{YukawaOmega}) to fix the Yukawa
couplings in terms of $\mbox{\boldmath$\Omega$}(z)$ given by one of the three
patterns (\ref{omegalep})-(\ref{omegadec}). Once $\mathbf{Y}_\nu(Q_0)$ is
determined by $\mbox{\boldmath$\Omega$}$, the RGEs for $\mathbf{Y}_\nu$, 
$M_A$ and $m_{\nu_B}$ can be used to calculate these quantities at any 
other scale
$M_1<Q<M_{\rm GUT}$. Indeed, one can obviously define the running of the 
coefficient of the ${\cal O}_5$ operator also for $Q\neq Q_0$, because this 
operator could be added to the initial Lagrangian with right-chiral neutrinos 
present. Above $Q_0$, the RGE for $\mathbf{C}$ is
\begin{eqnarray}
{d\over dt}\mathbf{C}=-K\mathbf{C}-
\left[(\mathbf{Y}_e^\dagger\mathbf{Y}_e)^T
+(\mathbf{Y}_\nu^\dagger\mathbf{Y}_\nu)^T\right]\mathbf{C}
-\mathbf{C}\left[(\mathbf{Y}_e^\dagger\mathbf{Y}_e)
+(\mathbf{Y}_\nu^\dagger\mathbf{Y}_\nu)\right],
\label{rge}
\end{eqnarray}
where $K=-6g^2_2-2g^2_y+2{\rm Tr}(3\mathbf{Y}_u^\dagger\mathbf{Y}_u+
\mathbf{Y}_\nu^\dagger\mathbf{Y}_\nu)$. By evolving $\mathbf{C}$ to an
arbitrary scale $Q$ by using the RGE (\ref{rge}) with $\mathbf{C}_0$ as 
the initial condition at $Q_0$~\footnote{For hierarchical light neutrino 
masses, the renormalization effects between the scale $M_Z$ and $Q_0$ on 
the mass eigenvalues and, in particular, on the mixing matrix $\Un{}$ 
are small, and $\mathbf{C}_0$ can be identified with 
$\Un{\ast}{\rm diag}(m_{\nu_1},m_{\nu_2},m_{\nu_3})\Un{\dagger}$
at the electroweak scale.} 
one can use (\ref{YukawaOmega}) to define $\tilde\OO{AB}$
at any scale $Q\neq Q_0$ by the formula
\begin{eqnarray}
\tilde\OO{AB}=-{i\over\sqrt{M_A}}
(\mathbf{O}_R\mathbf{Y}_\nu\mathbf{U}_\nu)^{AB}{1\over c_B},
\label{OmegaatQ}
\end{eqnarray}
where $\mathbf{U}_\nu$ diagonalizes $\mathbf{C}$ at $Q$, and 
$\mathbf{O}_R$ is the unitary 
matrix diagonalizing $M_R$ at $Q$. The question of consistency is now how 
much $\tilde\OO{AB}$ differs from $\OO{AB}$. In particular, $\tilde\OO{AB}(Q)$ 
defined in this way needs not be orthogonal, because in general 
$\mathbf{C}(Q)$ would be numerically different from
$\mathbf{Y}_\nu^T\mathbf{M}_R^{-1}\mathbf{Y}_\nu$ at the same scale.

However, it is easy to see that in supersymmetry $\tilde\OO{AB}(Q)$ defined
in this way {\it is} orthogonal because the RGE (\ref{rge}) is exactly the 
same
as the RGE of the combination 
$\mathbf{Y}_\nu^T\mathbf{M}^{-1}_R\mathbf{Y}_\nu$,
as obtained by using the chain differentiation rule and the RGEs of 
$\mathbf{Y}_\nu$ and $\mathbf{M}_R$ given, e.g., in~\cite{CHPO}. This is 
because, in
supersymmetry, the Yukawa couplings, the right-chiral 
neutrino mass term and the dimension-five operator are all $F$ terms.
Therefore, due to the non-renormalization theorems, the RG
running of their coefficients are entirely given by the wave function 
renormalization of the (super)fields out of which they are built.
It then follows that renormalization of the $N_A$ (super)fields,
which does not enter the running of $\mathbf{C}$, also cancels out in the 
expression for the RGE for $\mathbf{Y}_\nu^T\mathbf{M}^{-1}_R\mathbf{Y}_\nu$.

The RGE for $\tilde\OO{}$ can be then obtained
by simply differentiating the formula (\ref{OmegaatQ}), using the known 
RGEs for $\mathbf{U}_\nu$ and $c_B$ \cite{CHKRPO,CHPO}, the RGE for 
$\mathbf{Y}_\nu$ (see, e.g., the Appendix of~\cite{CHPO}~\footnote{In the 
expression (B.12) of this reference the factor of 3
should multiply $\mathbf{Y}_\nu^\dagger\mathbf{Y}_\nu$ instead of
$\mathbf{Y}_e^\dagger\mathbf{Y}_e$.}) and the RGEs for $M_A$ and 
$\mathbf{O}_R$ obtained by applying the technique 
of~\cite{BABU,CHKRPO,CHPO} to the RGE for 
$\mathbf{M_R}$ derived in~\cite{CAESIBNA}. One then gets
\begin{eqnarray}
&&{d\over dt}M_A=
4M_A\left[\mathbf{O}_R^\dagger(\mathbf{Y}_\nu\mathbf{Y}_\nu^\dagger)^T
\mathbf{O}_R\right]^{AA}
\phantom{aa}\nonumber\\
&&{d\over dt}\mathbf{O}^{AB}_R=\sum_C\mathbf{O}^{AC}_R\varepsilon_O^{CB}
\end{eqnarray}
where $\varepsilon_O^\dagger=-\varepsilon_O$, $\varepsilon_O^{AA}=0$ and
\begin{eqnarray}
\varepsilon_O^{AB}=-2{M_A+M_B\over M_A-M_B}{\rm Re}
\left[\mathbf{O}^\dagger_R(\mathbf{Y}_\nu\mathbf{Y}_\nu^\dagger)^T
\mathbf{O}_R\right]^{AB}
-2i{M_A-M_B\over M_A+M_B}{\rm Im}
\left[\mathbf{O}_R^\dagger(\mathbf{Y}_\nu\mathbf{Y}_\nu^\dagger)^T
\mathbf{O}_R\right]^{AB}
\end{eqnarray}
for $A\neq B$. Combining all the elements, one gets 
\begin{eqnarray}
{d\over dt}\tilde\OO{AB} = 
2{M_A\over\langle H\rangle^2}\sum_{D\neq A}\sum_C
\left[{M_Dm_{\nu_C}\over(M_D-M_A)}
\mathsf{Re}\left(\tilde\OO{AC}\tilde\OO{DC}^\ast\right)+
i{M_Dm_{\nu_C}\over(M_D+M_A)}
\mathsf{Im}\left(\tilde\OO{AC}\tilde\OO{DC}^\ast\right)\right]\tilde\OO{DB} 
\nonumber\\
+2{m_{\nu_B}\over\langle H\rangle^2}
\sum_{C\neq B}\sum_D\tilde\OO{AC}
\left[{m_{\nu_C}M_D\over(m_{\nu_B}-m_{\nu_C})}
\mathsf{Re}\left(\tilde\OO{DC}^\ast\tilde\OO{DB}\right)
+i{m_{\nu_C}M_D\over(m_{\nu_B}+m_{\nu_C})}
\mathsf{Im}\left(\tilde\OO{DC}^\ast\tilde\OO{DB}\right)\right]\nonumber\\
+2\sum_{C\neq B}\sum_D\tilde\OO{AC}y^2_{e_D}
\left[
{\sqrt{m_{\nu_B}m_{\nu_C}}\over m_{\nu_B}-m_{\nu_C}}
\mathsf{Re}\left(\mathbf{U}_\nu^{DC\ast}\mathbf{U}_\nu^{DB}\right)
+i{\sqrt{m_{\nu_B}m_{\nu_C}}\over m_{\nu_B}+m_{\nu_C}}
\mathsf{Im}\left(\mathbf{U}_\nu^{DC\ast}\mathbf{U}_\nu^{DB}\right)\right]
\phantom{aa}\label{rgeomega}
\end{eqnarray}
where the $y^2_{e_A}$ are the eigenvalues of $\mathbf{Y}_e^\dagger\mathbf{Y}_e$.

For non-degenerate heavy neutrinos, it is easy to see that for indices 
$AB$ 
corresponding to the small entries of $\tilde\OO{}$ in a given pattern 
(e.g., $\OO{1A}$ and $\OO{A1}$ with $A=2,3$ for the pattern 
(\ref{omegalep})), the first two lines are small compared to $1$ because 
they are always 
suppressed by the mass of the lightest left-chiral neutrino $m_{\nu_1}$. 
The only potentially dangerous contribution comes from $\tilde\OO{AC}=1$
in the last line. But this would require that $C=1$, and this term is 
suppressed 
at least by a factor $\sim\sqrt{m_{\nu_1}/m_{\nu_A}}y_\tau^2$, which 
is also small compared to $1$. Hence, for the patterns 
(\ref{omegalep})-(\ref{omegadec}) $\tilde\OO{}\approx\OO{}$ 
independently of the choice of $Q_0$, if the
heavy neutrinos are not degenerate.

Stability of the patterns (\ref{omegalep})-(\ref{omegadec}) may be more
problematic for arbitrarily tight degeneracy of the right-chiral 
neutrinos. One may, however, ask whether the pattern (\ref{omegadec}) 
is 
destabilized by the minimal degeneracy of $M_1$ and $M_2$ allowing for 
succesful leptogenesis with $M_1\simlt10^7$ GeV \cite{ELRAYA1,CHTU}. To 
make this estimate, we recall that with degenerate $M_1$ and $M_2$ the 
final lepton number asymmetry is roughly the same as would be obtained from 
decays of only one heavy neutrino having the effective mass 
$M_1^{\rm eff}=M_1/(M_2/M_1 -1)$. For the pattern (\ref{omegadec}), decays 
of 
a single neutrino produce the right lepton number asymetry if its mass is 
$\simgt10^{12}$ GeV \cite{CHTU}. Therefore, with the degeneracy we also need
$M_1^{\rm eff}\simgt10^{12}$~GeV. Consider now the contribution of the 
first 
line of (\ref{rgeomega}) to the derivative of the small element $\OO{B1}$, 
$B=1,2$. The most dangerous contribution for $A=1$ comes from $C=3$ and $D=2$.
It is of order $(M_1/\langle H\rangle^2)(M_2m_{\nu_3}/(M_2-M_1)\sim 
(M_1^{\rm eff}m_{\nu_3}/\langle H\rangle^2)\times\OO{21}\ll1$. Similarly the 
(nonresonant) contribution of $D=3$ is suppressed by small $\OO{33}$.
Therefore the pattern (\ref{omegadec}) is not destabilized in this case, 
either.


\end{document}